\documentclass[journal]{IEEEtran}
\usepackage{amsmath,amsfonts,amssymb,amsthm,epsfig}
\usepackage{graphicx}
\usepackage{caption}
\usepackage{subcaption}
\usepackage{enumerate}
\usepackage{hyperref}
\usepackage{url}
\usepackage{mathtools, cuted}

\usepackage{bbold}
\newtheorem{theorem}{Theorem}
\newtheorem{lemma}{\textit{Lemma}}
\newtheorem{corollary}{Corollary}
\newcommand\floor[1]{\lfloor#1\rfloor}

\begin{document}

\title{Classification and Representation via Separable Subspaces: Performance Limits and Algorithms}

\author{\IEEEauthorblockN{Ishan Jindal, \emph{Student Member, IEEE} and 
Matthew Nokleby, \emph{Member, IEEE}}\\
\IEEEauthorblockA{Department of Electrical and Computer Engineering,
Wayne State University,
Detroit, MI, 48202,  USA\\ Email: \{ishan.jindal, matthew.nokleby\}@wayne.edu}}

% \markboth{Journal of \LaTeX\ Class Files,~Vol.~14, No.~8, August~2015}%
% {Shell \MakeLowercase{\textit{et al.}}: Bare Demo of IEEEtran.cls for IEEE Journals}

\maketitle

\begin{abstract}
%Kronecker-structured (K-S) models recently have been proposed for the efficient representation, processing, and classification of multidimensional signals such as images and videos. Because they are tailored to the multi-dimensional structure of the target images, K-S models show improved performance in compression and reconstruction over more general (union of) subspace models. In this paper,
We study the classification performance of Kronecker-structured models in two asymptotic regimes and developed an algorithm for separable, fast and compact  K-S dictionary learning for better classification and representation of multidimensional signals by exploiting the structure in the signal. First, we study the classification performance in terms of {\em diversity order} and pairwise geometry of the subspaces. We derive an exact expression for the diversity order as a function of the signal and subspace dimensions of a K-S model. Next, we study the {\em classification capacity}, the maximum rate at which the number of classes can grow as the signal dimension goes to infinity. Then we describe a fast algorithm for \emph{Kronecker-Structured Learning of Discriminative Dictionaries} (\mbox{K-SLD$^2$}). Finally, we evaluate the empirical classification performance of K-S models for the synthetic data, showing that they agree with the diversity order analysis. We also evaluate the performance of \mbox{K-SLD$^2$} on synthetic and real-world datasets showing that the \mbox{K-SLD$^2$} balances compact signal representation and good classification performance.
\end{abstract}


% \fi
%\IEEEpeerreviewmaketitle

\section{Introduction} \label{sect::introduction}

The classification of high-dimensional signals arises in a variety of image processing settiings: object and digit recognition \cite{lee2005acquiring, bottou1994comparison}, speaker identification \cite{reynolds1995robust, kinnunen2010overview}, tumor classification \cite{ross2000systematic, soltani2015tensor}, and more. 
%[ISHAN: These last two references are pretty old. Is there nothing more recent on speaker identification or tumor classification? ANSWER: Added two reference both for speaker and tumor id ] 
A standard technique is to find a low-dimensional representation of the signal, such as a subspace or union of subspaces on which the signal approximately lies. However, for many signals, such as dynamic scene videos \cite{derpanis2012dynamic} or tomographic images \cite{tan2015tensor}, the signal inherently is multi-dimensional, involving dimensions of space and/or time. To use standard techniques, one vectorizes the signal, which throws out the spatial structure of the data which could be leveraged to improve representation fidelity, reconstruction error, or classification performance.

% via principal component analysis \cite{fisher1936use}, linear discriminant analysis, or sparse coding. Here, the main idea is to compactly represent the signal by a few coefficients in the overcomplete dictionary. For multi-dimensional signals: example includes dynamic scene videos \cite{derpanis2012dynamic}, tomographic images \cite{tan2015tensor}, where the signal of interest is not limited to two dimensional signal, in order to use the conventional matrix based approaches a high multi-dimensional signal is embed into a vector space by vectorizing the data. 
% %[ISHAN: Do these approaches vectorize the data and ignore the multi-dimensional structure?ANSWER: Yes These techniques are like simple neural networks, where we vectorize the given image and feed through the net.Addition: ] 
% For multi-dimensional signals, such as images and video, simply vectorizing the signal and applying sparse coding throws out the spatial structure of the data, which could be leveraged to improve representation fidelity and classification performance.

In order to exploit multi-dimensional signal structure, researchers have proposed {\em tensor-based} dictionary learning techniques, in which the signal of interest is a matrix or a higher-order tensor and the dictionary defining the (union of) subspace model is a tensor.
 A simple tensor-based model is the {\em Kronecker-structured} (K-S) model, in which a two-dimensional signal is represented by a coefficient matrix and two matrix dictionaries that pre- and post-multiply the coefficient matrix, respectively. Vectorizing this model leads to a dictionary that is the Kronecker product of two smaller dictionaries; hence the K-S model is a specialization of subspace models. This model is applied to spatio-temporal data in \cite{greenewald:CAMSAP13}, low-complexity methods for estimating K-S covariance matrices are developed in \cite{tsiligkaridis:TSP13}, and it is shown that the sample complexity of K-S models is smaller than standard union-of-subspace models in \cite{shakeri2016minimax}.

As standard union-of-subspace models have proven successful for classification tasks \cite{zhang2010discriminative,ramirez2010classification, jiang2011learning}, a natural question is the classification performance of K-S subspace models.
%Therefore, we expect that K-S union-of-subspace should do well for image classification tasks such as face recognition. These K-S subspaces can also be represneted using Gaussian mixture model (GMM) \cite{renna2016classification}, where each K-S subspace is associated with the covariance matrix of each Gaussian component. We use the standard subspace classification framework as discussed in \cite{ramirez2010classification} for Kronecker structured subspace classification [Long version]. We found that, indeed, K-S dictionaries are good at classification without any feature engineering and even more efficient than the standard union-of-subspace models. 
In this paper, we address this question from an information-theoretic perspective and developed an algorithm for learning discriminative K-S dictionaries. We consider a signal model in which each signal class is associated with a subspace whose basis is the Kronecker product of two smaller dictionaries; equivalently, we suppose that each signal class has a {\em matrix normal} distribution, where the row and column covariances are approximately low rank. Here the covariance of signal class follows a specific structure which is exactly the Kronecker product of two lower dimensional covariance matrices \cite{werner2008estimation, tsiligkaridis2013convergence, dutilleul1999mle}. In this sense, signals are drawn from a matrix Gaussian mixture model (GMM), similar to \cite{renna2016classification}, where each K-S subspace is associated with a mixture component.
%  in computer vision and signal processing tasks such as in image denoising and image restoration 

To find the underlying low dimensional representation of signals, dictionary learning methods are widely used \cite{yang2012coupled,bryt2008compression,elad2006image}. The underlying signal is compactly represented by a few large coefficients in an overcomplete dictionary. In a standard dictionary learning setting a 1-D signal $y_i$ is represented using a sparse coefficient vector $x_i$, where an overcomplete dictionary $D_i$ is learned by minimization problems similar to
\begin{equation}
\arg\,\min_{\{D_i,x_i\}} \sum_i||y_i -D_ix_i||_F^2 + \lambda ||x_i||_1
\label{eq::vectPre}
\end{equation}
Where $||\cdot||_F$ denotes the Forbenius norm, $||\cdot||_1$ denotes the $l_1$-norm, and $\lambda$ denotes the strength of the sparsity prior. Well-established methods for dictionary learning in this framework include K-SVD \cite{aharon2006img} and the method of optimal directions \cite{engan1999method}. These methods are targeted at dictionaries that faithfully {\em represent} the signal, and do not specifically consider classification.

Methods for incorporating discriminative ability into dictionary learning have been proposed, such as discriminative K-SVD \cite{zhang2010discriminative} (D-KSVD) and label consistent (LC-KSVD) \cite{jiang2013label}, which jointly learn a linear classifier and an overcomplete dictionary that is shared in common among the classes. Signals are then classified in the feature space induced by the dictionary. By contrast, \cite{yang2010metaface,ramirez2010classification,yang2011fisher,vu2017fast} propose methods for learning class-specific dictionaries, either by promoting incoherence among dictionaries or learning class-specific features. Signals are then classified by choosing the dictionary that minimizes the reconstruction error.

The above methods consider one-dimensional signals; multidimensional signals must first be vectorized, which may sacrifice structural information about the signal that could improve signal representation or classification. To preserve signal structure,\cite{duan2012k} extends K-SVD to tensor dictionaries, and \cite{zubair2013tensor, hawe2013separable,peng2014decomposable,soltani2015tensor} employ a variety of tensor decompositions to learn dictionaries tailored to multidimensional structure. These methods boast improved performance over traditional methods on a variety of signal processing tasks, including image reconstruction, image denoising and inpainting, video denoising, and speaker classification.

Similar to \cite{nokleby2015discrimination}, we first study the classification performance limits of K-S models in terms of {\em diversity order} and {\em classification capacity}, characterizing the performance in the limit of high SNR and large signal dimension, respectively. Further, we derive a tight upper bound on the misclassification probability in terms of the pairwise geometry of individual row and column subspaces. Where row and column subspaces correspond to two matrix dictionaries that pre- and post-multiply the coefficient matrix, respectively. We use principal angles between the subspaces as a measure to describe the geometry of subspaces \cite{huang2016role,hamm2008grassmann}. 

Finally, to learn discriminative dictionaries, we propose a new method, termed {\em Kronecker-Structured Learning of Discriminative Dictionaries} (\mbox{K-SLD$^2$}), that exploit multidimensional structure of the signal. \mbox{K-SLD$^2$} learns two subspace dictionaries per class: one to represent the columns of the signal, and one to represent the rows. Inspired by \cite{yang2011fisher}, we choose dictionaries that both represent each class individually and can be concatenated to form an overcomplete dictionary to represent signals generally. \mbox{K-SLD$^2$} is fast and learns compact data models with many fewer parameters than standard dictionary learning methods. We evaluate the performance of \mbox{K-SLD$^2$} on the Extended YaleB and UCI EEG database. The resulting dictionaries improve classification performance by up to 5\% when training sets are small, improve reconstruction performance across the board, and result in dictionaries with no more than  5\% of the storage requirements of existing subspace models.

% Also, inspired by \cite{yang2011fisher}, we propose a discriminative \emph{Kronecker-structured} (K-S) dictionary learning model for image classification by exploiting the structure in 2-D data. In which a 2-D signal is represented by a coefficient matrix and two smaller matrix dictionaries. In this approach we learn a set of dictionary pairs one pair for each class and enforce the class specific dictionary pair to well represent the corresponding class while the other dictionary pairs have poor representation ability for that class.

In Section \ref{sect::system.model}, we describe the K-S classification model in detail. In Section \ref{sect::Diversity.order} we derive the diversity order for K-S classification problems, showing the exponent of the probability of error as the SNR goes to infinity. This analysis depends on a novel expression, presented in Lemma \ref{lemma::FinalRank}, for the rank of sums of Kronecker products of tall matrices. In Section \ref{sect::Classification.capacity} we provide high-SNR approximations to the classification capacity. In Section \ref{sect::Dictionary.Learning}, we propose a discriminative K-S dictionary learning algorithm which balances the learning of class-specific, Kronecker-structured subspaces against the learning of an general overcomplete dictionary that allows for the representation of general signals.
% which learns a pair dictionary for each class and balances between the compact signal representation and good classification performance.
%\footnote{A long version of this paper, including complete proofs and extended numerical results, is available at \url{http://nokleby.eng.wayne.edu/papers/isit.long.pdf}.}
 In Section \ref{sect::Numerical.Result} we show that the empirical classification performance of K-S models agrees with the diversity analysis and evaluate the performance of proposed discriminative algorithm on extended YaleB face recognition dataset and EEG signal dataset correlating the EEG signals with individual's alcoholism.

%We observe that, for this Kronecker structured model we achieve the same misclassification error as for subspace model. Kronecker structured model are less complex than subspace model. 
%We provided a very general formula for rank of a sum of two Kronecker products. 
%
%Dimensionality reduced representation for classification\\
%balances dimensionality and expressiveness\\
%linear subspace models are common\\
%kronecker structure emerging for representation\\
%This paper: analysis of KS models for classsification\\
%contributions:\\
%bounds on classification capacity\\
%expression for the diversity gain\\
%numerical results: synthetic and YaleB

\section{Problem Definition} \label{sect::system.model}

\subsection{Kronecker-structured Signal Model}
To formalize the classification problem, let the signal of interest $\mathbf{Y} \in \mathbb{R}^{m_1\times m_2}$ be a matrix whose entries are distributed according to one of $L$ class-conditional densities $p_l(\mathbf{Y})$. Each class-conditional density corresponds to a Kronecker-structured model described by the pair of matrices $\mathbf{A}_l \in \mathbb{R}^{m_1 \times n_1}$ and $\mathbf{B}_l \in \mathbb{R}^{m_2 \times n_2}$. The matrix $\mathbf{A}_l$ describes the subspace on which the columns of $\mathbf{Y}$ approximately lie, and $\mathbf{B}_l$ describes the subspace on which the rows of $\mathbf{Y}$ approximately lie. More precisely, if $\mathbf{Y}$ belongs to class $l$, it has the form
\begin{equation}
	\mathbf{Y} = \mathbf{A}_l \mathbf{X} \mathbf{B}_l^T + \mathbf{Z},
\label{eq::model}
\end{equation}
where $\mathbf{Z} \in \mathbb{R}^{m_1 \times m_2} $ has i.i.d. zero-mean Gaussian entries with variance $\sigma^2 > 0$, and $\mathbf{X} \in \mathbb{R}^{n_1 \times n_2}$ has i.i.d. zero-mean Gaussian entries with unit variance. We can also express $\mathbf{Y}$ in vectorized form:
%which describes the deviation of the true signal from its postulated Kronecker structured model. Here, $\mathbf{A} \in \mathbb{R}^{m_1 \times n_1}$ and $\mathbf{B}\in \mathbb{R}^{m_2 \times n_2}$ containes the basis of the row and column subspace respectively. We can vectorize \eqref{eq::model} using a Kronecker product property \cite{van2000ubiquitous}, $\mathrm{vec}(\mathbf{A}\mathbf{X}\mathbf{B}^T) = (\mathbf{B}\otimes \mathbf{A})\mathrm{vec}(\mathbf{X})$ to obtain the expression for observation vector $\mathbf{y} = \mathrm{vec}(\mathbf{Y}) \, \in \mathbb{R}^{M}$:
\begin{equation}
\mathbf{y} = (\mathbf{B}_l \otimes \mathbf{A}_l)\mathbf{x} + \mathbf{z}, \label{eqn:class.conditional.density}
\end{equation}
for coefficient vector $\mathbf{x} = \mathrm{vec}(\mathbf{X}) \, \in \mathbb{R}^{N}$, and noise vector $\mathbf{z} \in \mathbb{R}^{M}$, where $N = n_1n_2$, $M = m_1m_2$, and where $\otimes$ is the usual Kronecker product. Then, the class-conditional density of $\mathbf{y}$ is
\begin{equation}
	p_l(\mathbf{y}) = \mathcal{N}(0, (\mathbf{B}_l \otimes \mathbf{A}_l)(\mathbf{B_l}\otimes \mathbf{A_l})^T + \sigma^2 \cdot \mathbf{I}).
\end{equation}
In other words, the vectorized signal $\mathbf{y}$ lies near a subspace with a Kronecker structure that encodes the row and column subspaces of $\mathbf{Y}$.

In the sequel, we will characterize the performance limits over ensembles of classification problems of this form. %To this end, define the set 
% \begin{multline}
% 	\mathcal{Q}(m_1,m_2,n_1,n_2) = \{ \mathcal{N}(0, (\mathbf{B} \otimes \mathbf{A})(\mathbf{B} \otimes \mathbf{A})^T) : \\ \mathbf{A} \in \mathbb{R}^{m_1 \times n_1}, \mathbf{B}\in \mathbb{R}^{m_2 \times n_2} \}
% \end{multline}
%of class-conditional densities having the form of $p_l(\mathbf{y})$. 
To this end, we parameterize the set of class-conditional densities via
\begin{equation}
	\mathcal{A}(m_1,m_2,n_1,n_2) = \mathbb{R}^{m_1 \times n_1} \times \mathbb{R}^{m_2 \times n_2},
\end{equation}
which contains the set of matrices indicating the row and column subspaces given signal and subspace dimensions $m_1,m_2,n_1,n_2$. We can represent an $L$-ary classification problem by a tuple $\mathbf{a} = (a_1,\cdots, a_L) \in \mathcal{A}^L(m_1,m_2,n_1,n_2)$, where each $a_l \in \mathcal{A}(m_1,m_2,n_1,n_2)$ is the pair of matrices $a_i = (\mathbf{A}_l, \mathbf{B}_l)$.
%We parameterize the set $\mathcal{Q}(m_1,m_2,n_1,n_2)$ by $\mathcal{A}(m_1,m_2,n_1,n_2) = \mathbb{R}^{m_1 \times n_1} \times \mathbb{R}^{m_2 \times n_2} $ is the Cartesian product of row and column subspaces. \\ We can represent a tensor classification problem by defining a tuple $ \mathbf{a} = (a_1,\cdots, a_L) \in \mathcal{A}^L(m_1,m_2,n_1,n_2)$, where each $a_i \in \mathcal{A}(m_1,m_2,n_1,n_2)$ is the pair of row and column subspaces that is $a_i = (\mathbf{A}_i, \mathbf{B}_i)$.
% where $\mathcal{A}^L(m_1,m_2,n_1,n_2)$ is the $L$-fold Cartesian product of $\mathcal{A}(m_1,m_2,n_1,n_2)$
Let $ p(\mathbf{y} |a_l) = p(\mathbf{y} | \mathbf{A}_l,\mathbf{B}_l) = p_l(\mathbf{y})$, for $1 \leq l \leq L$, denote the class conditional densities parametrized by $\mathbf{a}\in \mathcal{A}(m_1,m_2,n_1,n_2)$.
%We parameterize the set $\mathcal{Q}(m_1,m_2,n_1,n_2)$ by $\mathbf{A_{row}}(m_1,n_1) = \mathbb{R}^{m_1 \times n_1}$ and $\mathbf{\mathbf{B}_{column}}(m_2,n_2) = \mathbb{R}^{m_2 \times n_2} $. \\ By defining tuples $ a = (a_1,\cdots, a_L) \in \mathbf{A^L_{row}}(m_1,n_1)$ and $b = (b_1,\cdots, b_L) \in \mathbf{B^L_{colun}}(m_2,n_2)$, where $\mathbf{A^L_{row}}(m_1,n_1)$ and $\mathbf{B^L_{colun}}(m_2,n_2)$ are the $L$-fold Cartesian product of $\mathbf{A_{row}}(m_1,n_1)$ and $\mathbf{B_{column}}(m_2,n_2)$ respectively, we can represent a tensor classification problem by  $c = (\{a_1,b_1\},\cdots,\{a_L,b_L\})\in \mathbf{C}^L(m_1m_2,n_1n_2) $, where $\mathbf{C}^L(m_1m_2,n_1n_2) $ is the $L$ number of $(a,b)$ tuple pair. Let $p(\mathbf{X},c_l) = p_l(\mathbf{X})$, for $1 \leq l \leq L$, denotes the class conditional densities parametrized by $a\in \mathbf{A_{row}}(m_1,n_1)$ and $ b \in \mathbf{B_{column}}(m_2,n_2)$.
For a classification problem defined by $\mathbf{a}$, we can define the average misclassification probability:
\begin{equation}
	P_e(\mathbf{a}) = \frac{1}{L}\sum_{l=1}^L \Pr(\hat{l} \neq l | \mathbf{y} \sim p(\mathbf{y} | a_l),
\end{equation}
where $\hat{l}$ is the output of the maximum-likelihood classifier over the class-conditional densities described by $\mathbf{a}_l$. In this paper, we provide two asymptotic analyses of $P_e(\mathbf{a})$. First, we consider the {\em diversity order}, which characterizes the slope of $P_e(\mathbf{a})$ for a particular $\mathbf{a}$ as $\sigma^2 \to 0$. Second, we consider the {\em classification capacity}, which characterizes the asymptotic error performance {\em averaged over } $\mathbf{a}$ as $ n_1,m_1,n_2,m_2$ go to infinity. For the latter case, we define a prior distribution over the matrix pairs $(\mathbf{A}_l,\mathbf{B}_l)$ in each class:
\begin{equation}
	p(\mathbf{a}) = \prod_{p=1}^{m_1}\prod_{q=1}^{n_1} \prod_{r=1}^{m_2}\prod_{s=1}^{n_2}  \mathcal{N}(a_{pq};0,1/n_1) \cdot \mathcal{N}(b_{rs};0,1/n_2) \label{eqn:model.prior}
\end{equation}
where $a_{pq}$ is the $(p,q)$th element of matrix $\mathbf{A}$ and $b_{rs}$ is the $(r,s)$th element of matrix $\mathbf{B}$. 
%[ISHAN: This is wrong! The entries of $B \otimes A$ are NOT i.i.d.!]
Note that the column and row subspaces described by $\mathbf{A}$ and $\mathbf{B}$ are uniformly distributed over the Grassmann manifold because the matrix elements are i.i.d. Gaussian; however, the resulting K-S subspaces are not uniformly distributed.

\subsection{Diversity Order}
For a fixed classification problem $\mathbf{a}$, the diversity order characterizes the decay of the misclassification probability as the noise power goes to zero. By analogy with the definition of the diversity order in wireless communications \cite{zheng2002communication}, we consider the asymptotic slope of $P_e(\mathbf{a})$ on a logarithmic scale as $\sigma^2 \to 0$ that is the mismatch between data and model is vanishingly small. Formally, the diversity order is defined as
\begin{equation}
	d(\mathbf{a}) = \lim_{\sigma^2 \rightarrow 0}-\frac{\log P_e(\mathbf{a})}{\frac{1}{2}\log(1/\sigma^2)} \label{eq::model::8}.
\end{equation}
In Section \ref{sect::Diversity.order}, we characterize exactly the diversity order for almost every $\mathbf{a}$.

\subsection{Classification Capacity}
The classification capacity characterizes the number of unique subspaces that can be discerned as $n_1$, $n_2$, $m_1$ and $m_2$ go to infinity. That is, we derive bounds on how fast the number of classes $L$ can grow as a function of signal dimension while ensuring the misclassification probability decays to zero almost surely. Here, we define a variable $m$\footnote{Note that $m$ is different from $M$, where $m$ is the variable we let to go to infinity and $M = m_1m_2$.} and let it go to infinity. As $m$ grows to infinity we let the dimensions $m_1$, $m_2$, $n_1$ and $n_2$ scale linearly with $m$ as follows:
\begin{multline}
m_1(m) = \floor{\kappa_1 m}, \, m_2(m) = \floor{\kappa_2 m}, \\n_1(m) = \floor{\upsilon_1 m}, n_2(m) = \floor{\upsilon_2 m}
\end{multline}
for $ \upsilon_1, \upsilon_2 \geq 1$ and $ 0 \leq \kappa_1, \kappa_2 \leq 1$. We let the number of classes $L$ grow exponentially in $m$ as:
\begin{equation}
	L(m) = \floor{2^{\rho m_1(m)m_2(m)}},
\end{equation}
for some $\rho \geq 0$, which we call the {\em classification rate}. We say that the classification rate $\rho$ is achievable if $\lim_{m \rightarrow \infty} E[P_e(\mathbf{a})] = 0$.
% \begin{multline}
% 	\lim_{m \rightarrow \infty} E[P_e(\mathbf{a})] = \lim_{m \rightarrow \infty}\\ \int_{\mathcal{A}^{L(m)}(m_1(m),m_2(m),n_1(m)n_2(m))} P_e(\mathbf{a}) \prod_{i=1}^{L(m)}p(a_i)da = 0.
% \end{multline} 
For fixed signal dimension ratios $\upsilon_1,\upsilon_2,\kappa_1$ and $\kappa_2$, we define $C(\upsilon_1,\upsilon_2,\kappa_1,\kappa_2)$ as the supremum over all achievable classification rates, and we call $C(\upsilon_1,\upsilon_2,\kappa_1,\kappa_2)$ (sometimes abbreviated by $C$) the {\em classification capacity}.

We can bound the classification capacity by the mutual information between the signal vector $\mathbf{y}$ and the matrix pair $(\mathbf{A}$, $\mathbf{B})$ that characterizes each Kronecker-structured class.
\begin{lemma} 
\label{lemma::classcapacity}
	The classification capacity satisfies:
	\begin{equation}
		C \leq \lim_{m \rightarrow \infty} \frac{I(\mathbf{y};\mathbf{A},\mathbf{B})}{m_1(m)m_2(m)}
	\label{eq::model::capacity}
	\end{equation}
	Where the mutual information is computed with respect to $p(\mathbf{a})$.
\end{lemma}
% \begin{proof}
% By Fano's inequality \cite{cover2012elements}, we obtain $$P_e(\mathbf{a}) \geq 1-\frac{I(\mathbf{y};\mathbf{A},\mathbf{B}) - 1}{m_1(m)m_2(m)\rho}$$ which is bounded away from zero when $\rho$ exceeds the RHS of \eqref{eq::model::capacity}.
% \end{proof}

To prove lower bounds on the diversity order and classification capacity, we will need the following lemma, which gives the well-known Bhattacharyya bound on the probability of error of a maximum-likelihood classifier that chooses between two Gaussian hypotheses.
\begin{lemma}[\cite{kailath1967divergence}]
	\label{lemma::Bhattacharyabpund}
 	Consider a signal distributed according to $\mathcal{N}(\mu_1, \Sigma_1)$ or $\mathcal{N}(\mu_2, \Sigma_2)$ with equal priors. Then, define
	\begin{equation}
	b = \frac{1}{2} \ln\left(\frac{|\frac{\Sigma_1+\Sigma_2}{2}|}{|\Sigma_1|^{\frac{1}{2}}|\Sigma_2|^{\frac{1}{2}}}\right) + \frac{1}{8}(\mu_1-\mu_2)\left[\frac{\Sigma_1+\Sigma_2}{2}\right]^{-1}(\mu_1-\mu_2)
\end{equation}
Supposing maximum likelihood classification, the misclassification probability is bounded by 
\begin{equation}
	P_e(\mu_1,\Sigma_1,\mu_2,\Sigma_2) \leq \frac{1}{2}\exp(-b).
\label{eq::model::7}
\end{equation}
\end{lemma}

\subsection{Subspace Geometry}
% Kronecker subspaces are the subset of linear subspaces and can be re

We characterize the subspace geometry in terms of \emph{principal angles}. Principal angle defines as the canonical angles between elements of subspaces, and they induce a distance metric on the Grassmann manifold. If the principal angles between subspaces is large, this means that the subspaces are far apart and easily discernible.

Consider two linear subspaces $\mathcal{A}_1$ and $\mathcal{A}_2$ of $\mathbb{R}^{m}$ with same dimensions $n$ each. The \emph{principal angles} between these two subspaces are defined recursively as follows:
\begin{align*}
\cos(\theta_t) &= \max_{u_t \in \mathcal{A}_1} \max_{v_t \in \mathcal{A}_2} u_t^T v_t \\ &  \, \text{subject to } \quad u_t^T u_t = 1,\, v_t^T v_t = 1,\\
&\,\, \quad\quad \quad \quad \quad u_t^T u_i = 0, \,v_t^T v_t = 0, \quad (i < t)
\end{align*}
where $0\leq \theta_1 \leq \theta_2 \leq \cdots \leq \theta_{n_1} \leq \frac{\pi}{2}$ and the first principal angle $\theta_1$ is the smallest angle between all pairs of unit vectors in the first and the second subspaces \cite{knyazev2012principal}. 

The principal angles can be computed directly via computing the \emph{singular value decomposition (SVD)} of $\mathbf{A}_1^T\mathbf{A}_2$, where $\mathbf{A}_1$ and $\mathbf{A}_2$ are orthonormal basis for the subspaces $\mathcal{A}_1$ and $\mathcal{A}_2$, respectively.

\begin{equation*}
\mathbf{A}_1^T\mathbf{A}_2 = \mathbf{U}_A\cos(\Theta_A)\mathbf{V}_A^T,
\end{equation*}
where the cosine of principal angles, $\cos(\Theta_A) = \text{diag}(\cos(\theta_1^A), \cos(\theta_2^A), \cdots, \cos(\theta_{n_1}^A))$, are the singular values of $\mathbf{A}_1^T\mathbf{A}_2$.

In this problem, suppose $\mathbf{A}_1$ and $\mathbf{A}_2$ are orthonormal basis for the  subspaces $\mathcal{A}_1$ and $\mathcal{A}_2$ on which columns of signal approximately lies and $\mathbf{B}_1$ and $\mathbf{B}_2$ are orthonormal basis for the  subspaces $\mathcal{B}_1$ and $\mathcal{B}_2$ on which rows of signal approximately lies. Then we define the orthonormal basis $\mathbf{D}_1 = \mathbf{B}_1 \otimes \mathbf{A}_1$ and $\mathbf{D}_2 = \mathbf{B}_2  \otimes \mathbf{A}_2$ for the Kronecker-structured subspaces $\mathcal{D}_1$ and $\mathcal{D}_2$, respectively. The cosine of principal angles between $\mathbf{D}_1$ and $\mathbf{D}_2$ are the singular values of $\mathbf{D}_1^T\mathbf{D}_2$ as follows:
\begin{align*}
\mathbf{D}_1^T\mathbf{D}_2 &= (\mathbf{B}_1 \otimes \mathbf{A}_1)^T(\mathbf{B}_2 \otimes \mathbf{A}_2)\\
% &=(\mathbf{B}_1^T \otimes \mathbf{A}_1^T)(\mathbf{B}_2 \otimes \mathbf{A}_2)\\
&= (\mathbf{B}_1^T\mathbf{B}_2)  \otimes (\mathbf{A}_1^T\mathbf{A}_2)\\
&= (\mathbf{U}_B\cos(\Theta_B)\mathbf{V}_B^T) \otimes (\mathbf{U}_A\cos(\Theta_A)\mathbf{V}_A^T)\\
% &= (\mathbf{U}_B\otimes \mathbf{U}_A  ) ( \cos(\Theta_B) \otimes \cos(\Theta_A)) (\mathbf{V}_B^T\otimes  \mathbf{V}_A^T )\\
&= (\mathbf{U}_B \otimes \mathbf{U}_A  ) (\cos(\Theta_B) \otimes \cos(\Theta_A)) ( \mathbf{V}_B \otimes \mathbf{V}_A )^T\\
&= \mathbf{U} \cos(\Theta) \mathbf{V}^T,
\end{align*}
where the cosine of principal angle between two Kronecker subspaces is the Kronecker product of cosine of principal angles between two row subspaces and two column subspaces that is $\cos(\Theta) = \cos(\Theta_A)\otimes \cos(\Theta_B)$.

\section{Diversity Order} \label{sect::Diversity.order}

As mentioned in Section \ref{sect::system.model}, the diversity order measures how quickly misclassification probability decays with the noise power for a fixed number of discernible subspaces. By careful analysis   using the Bhattacharrya bound, we derive an exact expression for the diversity order for almost every\footnote{With respect to the Lebesgue measure over $\mathcal{A}^L$.} classification problem. First, we state an expression that holds in general.
\begin{theorem}
	For a classification problem described by the tuple $\mathbf{a} \in \mathcal{A}^L$ such that $r(\mathbf{A}_l)=n_1$ and $r(\mathbf{B}_l)=n_2$ for every $l$, the diversity order is $d(\mathbf{a}) = r^* - n_1n_2$, where
	\begin{equation}
		r^* = \min_{i,j} r(\begin{bmatrix} \mathbf{B}_i \otimes \mathbf{A}_i & \mathbf{B}_j \otimes \mathbf{A}_j \end{bmatrix}),
	\end{equation}
	and where $r(\cdot)$ denotes the matrix rank.
	\label{thm:diversity.order}
\end{theorem}
\begin{IEEEproof}
Applying the Bhattacharyya bound, the probability of a {\em pairwise error} between two Kronecker-structured classes $i$ and $j$ with covariances $$\Sigma_i = \mathbf{D}_i\mathbf{D}_i^T + \sigma^2 \mathbf{I}, \quad \Sigma_j = \mathbf{D}_j\mathbf{D}_j^T + \sigma^2 \mathbf{I},$$ is bounded by 
\begin{equation}
P_e(\mathbf{D}_i,\mathbf{D}_j) \leq \frac{1}{2} \left(\frac{|\frac{\mathbf{D}_i\mathbf{D}_i^T + \mathbf{D}_j\mathbf{D}_j^T + 2\sigma ^2 \mathbf{I}}{2}|}{|\mathbf{D}_i\mathbf{D}_i^T + \sigma ^2 \mathbf{I}|^{\frac{1}{2}}|\mathbf{D}_j\mathbf{D}_j^T + \sigma ^2 \mathbf{I}|^{\frac{1}{2}}} \right)^{-\frac{1}{2}}
\label{eq::Bhattacharya}
\end{equation}
where
\begin{align*}
	\mathbf{D}_i\mathbf{D}_i^T &= \mathbf{B}_i\mathbf{B}_i^T \otimes \mathbf{A}_i\mathbf{A}_i^T, \\
	\mathbf{D}_j\mathbf{D}_j^T &= \mathbf{B}_j\mathbf{B}_j^T \otimes \mathbf{A}_j\mathbf{A}_j^T.
\end{align*}
%With probability one, the matrices $ \mathbf{D}_j\mathbf{D}_j^T$ and $\mathbf{D}_i\mathbf{D}_i^T$ have rank $n_1n_2$. Expanding the matrix $(\mathbf{D}_i\mathbf{D}_i^T+\mathbf{D}_j\mathbf{D}_j^T)$ as
% \begin{align}
% \mathbf{D}_i\mathbf{D}_i^T+\mathbf{D}_j\mathbf{D}_j^T &= \mathbf{B}_i \mathbf{B}_i ^T \otimes \mathbf{A}_i \mathbf{A}_i ^T + \mathbf{B}_j \mathbf{B}_j ^T \otimes \mathbf{A}_j \mathbf{A}_j ^T
% \label{eq::CCB::4} 
% \end{align}
Using the well-known Kronecker product identities $ (p \otimes q)\cdot(r \otimes s) = (pq \otimes rs) $ and $(p \otimes r)^T = (p^T \otimes r^T)  $ we can write the matrix 
$\mathbf{D}_i\mathbf{D}_i^T + \mathbf{D}_j\mathbf{D}_j^T$ as
\begin{equation}
\mathbf{D}_i\mathbf{D}_i^T+\mathbf{D}_j\mathbf{D}_j^T = \begin{bmatrix}
\mathbf{B}_i \otimes \mathbf{A}_i & \mathbf{B}_j \otimes \mathbf{A}_j
\end{bmatrix} \cdot \begin{bmatrix}
\mathbf{B}_i ^T \otimes \mathbf{A}_i ^T \\ \mathbf{B}_j ^T \otimes \mathbf{A}_j ^T
\end{bmatrix}
\end{equation}
It is trivial that $r(A) = r(AA^T)$, thus
\begin{equation*}
r(\mathbf{D}_i\mathbf{D}_i^T+\mathbf{D}_j\mathbf{D}_j^T ) = r(\begin{bmatrix}
\mathbf{B}_i \otimes \mathbf{A}_i & \mathbf{B}_j \otimes \mathbf{A}_j
\end{bmatrix}) = r_{ij}^*
\end{equation*}
Let $\lambda_{i}$ and $\lambda_{j}$ denote the nonzero eigenvalues of $\mathbf{D}_i\mathbf{D}_i^T$ and $\mathbf{D}_j\mathbf{D}_j^T$ respectively, and let $\lambda_{ij}$ denote the nonzero eigenvalues of $\mathbf{D}_i\mathbf{D}_i^T + \mathbf{D}_j\mathbf{D}_j^T $ and $r^*_{ij}$ denote its rank. Then, we can write the pairwise bound in \eqref{eq::CCB::13}.
% \begin{align}
% \label{eq::CCB::13}
% P_e(\mathbf{D}_i,\mathbf{D}_j) &\leq \frac{1}{2} \left( \frac{(\sigma^2)^{m_1m_2- r^*_{ij}}\prod_{l=1}^{r^*_{ij}}(\lambda_{ijl}+\sigma^2)}{\sqrt{(\sigma^2)^{m_1m_2-n_1n_2}\prod_{l=1}^{n_1n_2}(\lambda_{il} + \sigma^2) \cdot (\sigma^2)^{m_1m_2-n_1n_2}\prod_{l=1}^{n_1n_2}(\lambda_{jl}+\sigma^2)}} \right)^{-\frac{1}{2}}\\
% \label{eq::CCB::14}{}
% & = \frac{1}{2} \left(\frac{1}{\sigma^2}\right)^{-\frac{r^*_{ij}-n_1n_2}{2}} \cdot \left( \frac{\prod_{l=1}^{r^*_{ij}}(\lambda_{ijl}+\sigma^2)}{\sqrt{\prod_{l=1}^{n_1n_2}(\lambda_{il} + \sigma^2) \cdot \prod_{l=1}^{n_1n_2}(\lambda_{jl}+\sigma^2)}} \right)^{-\frac{1}{2}}
% \end{align}
\begin{figure*}[!htbp]
\begin{align}
\label{eq::CCB::13}
P_e(\mathbf{D}_i,\mathbf{D}_j) &\leq \frac{1}{2} \left( \frac{(\sigma^2)^{m_1m_2- r^*_{ij}}\prod_{l=1}^{r^*_{ij}}(\lambda_{ijl}+\sigma^2)}{\sqrt{(\sigma^2)^{m_1m_2-n_1n_2}\prod_{l=1}^{n_1n_2}(\lambda_{il} + \sigma^2) \cdot (\sigma^2)^{m_1m_2-n_1n_2}\prod_{l=1}^{n_1n_2}(\lambda_{jl}+\sigma^2)}} \right)^{-\frac{1}{2}}\\
\label{eq::CCB::14}{}
& = \frac{1}{2} \left(\frac{1}{\sigma^2}\right)^{-\frac{r^*_{ij}-n_1n_2}{2}} \cdot \left( \frac{\prod_{l=1}^{r^*_{ij}}(\lambda_{ijl}+\sigma^2)}{\sqrt{\prod_{l=1}^{n_1n_2}(\lambda_{il} + \sigma^2) \cdot \prod_{l=1}^{n_1n_2}(\lambda_{jl}+\sigma^2)}} \right)^{-\frac{1}{2}}
\end{align}
\hrulefill
\end{figure*}
%\footnotetext{Under no conditions rank of sum of Kronecker product achieves $m_1m_2$}
By construction,$$ \mathbf{D}_i\mathbf{D}_i^T+\mathbf{D}_j\mathbf{D}_j^T \geq  \mathbf{D}_i\mathbf{D}_i^T, \mathbf{D}_j\mathbf{D}_j^T $$
Using Weyl's monotonicity theorem $2\lambda_{ijl} \geq \lambda_{il}$ and $2\lambda_{ijl} \geq \lambda_{jl}$ for every $ 1\leq l \leq n_1n_2$, Therefore,
 \begin{align*}
 \prod_{l=1}^{n_1n_2}2(\lambda_{ijl} + \sigma^2) \geq \sqrt{\prod_{l=1}^{n_1n_2}(\lambda_{il} + \sigma^2) \cdot \prod_{l=1}^{n_1n_2}(\lambda_{jl}+\sigma^2)}
 \end{align*}
From this we can write
 \begin{align}
 P_e(\mathbf{D}_i,\mathbf{D}_j) &\leq  \frac{1}{2} \left(\frac{1}{\sigma^2}\right)^{-\frac{r^*_{ij}-n_1n_2}{2}} \cdot 2^{\frac{n_1n_2}{2}} \nonumber \\ & \quad \cdot  \left( \prod_{l=n_1n_2+1}^{r^*_{ij}}(\lambda_{ijl}+\sigma^2) \right)^{-\frac{1}{2}} \\
 \label{eq::CCB::10}
& \leq 2^{\frac{n_1n_2-2}{2}} \left(\frac{1}{\sigma^2}\right)^{-\frac{r^*_{ij}-n_1n_2}{2}} \nonumber \\ & \quad
 \cdot  (\lambda_{ijr^*_{ij}}+\sigma^2)^{-\frac{r^*_{ij}-n_1n_2}{2}}\\
 \label{eq::CCB::9} 
 & = 2^{\frac{n_1n_2-2}{2}} \left( 1+\frac{\lambda_{ijr^*_{ij}}}{\sigma^2}\right)^{-\frac{r^*_{ij}-n_1n_2}{2}}
 \end{align}
 Next, we bound $P_e(\mathbf{a}) \leq \sum_{i \neq j}P_e(\mathbf{D}_i,\mathbf{D}_j)$ via the union bound. 
For all the $L$ subspaces, we obtain the pairwise error probability and by invoking the union bound over all the subspaces we obtain:
 \begin{align*}
 P_e(\mathbf{a}) &\leq \frac{1}{L}\sum_{l=1}^{L}\sum_{l\neq\hat{l}}P_e(\mathbf{D}_l,\mathbf{D}_{\hat{l}})\\
 &= (L-1)P_e(\mathbf{D}_l,\mathbf{D}_{\hat{l}})\\
 &\leq 2^{\rho m_1m_2} P_e(\mathbf{D}_l,\mathbf{D}_{\hat{l}})
 \end{align*}
 Taking logarithm on both sides we obtain:
 \begin{multline}
 \log_2(P_e(\mathbf{a})) \leq \rho m_1m_2 +\frac{n_1n_2-2}{2}- \\ \frac{r^*_{ij}-n_1n_2}{2}\log_2\left( 1+\frac{\lambda_{ijr^*_{ij}}}{\sigma^2}\right)
 \label{eq::CCB::11} 
 \end{multline}
% Next, we bound $P_e(\mathbf{a}) \leq \sum_{i \neq j}P_e(\mathbf{D}_i,\mathbf{D}_j)$ via the union bound. 
% For all the $L(m)$ subspaces, we obtain the pairwise error probability and by invoking the union bound over all the subspaces we obtain:
%  \begin{align*}
%  E[P_e(\mathbf{a})] &\leq \frac{1}{L(m)}\sum_{l=1}^{L(m)}\sum_{l\neq\hat{l}}E[P_e(\mathbf{D}_l,\mathbf{D}_{\hat{l}})]\\
%  &= (L(m)-1)E[P_e(\mathbf{D}_l,\mathbf{D}_{\hat{l}})]\\
%  &\leq 2^{\rho m_1m_2} E[P_e(\mathbf{D}_l,\mathbf{D}_{\hat{l}})]
%  \end{align*}
%  Taking logarithm on both sides we obtain:
%  \begin{multline}
%  \log_2(E[P_e(\mathbf{a})]) \leq \rho m_1m_2 +\frac{n_1n_2-2}{2}- \\ \frac{r^*_{ij}-n_1n_2}{2}\log_2\left( 1+\frac{\lambda_{ijr^*_{ij}}}{\sigma^2}\right)
%  \label{eq::CCB::11} 
%  \end{multline}
Putting this and (\ref{eq::CCB::9}) into the definition of the diversity order from (\ref{eq::model::8}), we obtain
\begin{align}
	d(\mathbf{a}) &\geq \min_{i,j} \lim_{\sigma^2 \to 0} -\frac{-\frac{r^*_{ij} - n_1n_2}{2}\log(1/\sigma^2)}{\frac{1}{2}\log(1/\sigma^2)} \\
	&= \min_{i,j} r^*_{ij} - n_1n_2 \\
	&= r^* - n_1n_2.
\end{align}
Finally, \cite{kailath1967divergence} shows that the Bhattacharyya bound is exponentially tight as the pairwise error decays to zero. Furthermore, the union bound is exponentially tight. Therefore, the above inequality holds with equality, and $d(\mathbf{a})~=~r^*~-~n_1n_2.$ 
% \begin{multline}
% P_e(\mathbf{D}_i,\mathbf{D}_j) \leq  \frac{1}{2} \left(\frac{1}{\sigma^2}\right)^{-\frac{r^*-n_1n_2}{2}} \cdot 2^{\frac{n_1n_2}{2}} \cdot \\ \left( \prod_{l=n_1n_2+1}^{r^*}(\lambda_{ijl}+\sigma^2) \right)^{-\frac{1}{2}} 
% \end{multline}
% \begin{multline}
% \label{eq::CCB::10}
% P_e(\mathbf{D}_i,\mathbf{D}_j) \leq 2^{\frac{n_1n_2-2}{2}} \left(\frac{1}{\sigma^2}\right)^{-\frac{r^*-n_1n_2}{2}}
% \cdot \\ (\lambda_{ijr^*}+\sigma^2)^{-\frac{r^*-n_1n_2}{2}}
% \end{multline}
% \begin{multline}
% \label{eq::CCB::9} 
% P_e(\mathbf{D}_i,\mathbf{D}_j) \leq 2^{\frac{n_1n_2-2}{2}} \left( 1+\frac{\lambda_{ijr^*}}{\sigma^2}\right)^{-\frac{r^*-n_1n_2}{2}}
% \end{multline}
\end{IEEEproof}

For almost every classification problem, the rank $r^*$ has the same value, as we show in the next lemma.
\begin{lemma} 
\label{lemma::FinalRank}
For almost every classification problem $\mathbf{a}$, the matrices $\begin{bmatrix}
\mathbf{B}_i \otimes \mathbf{A}_i & \mathbf{B}_j \otimes \mathbf{A}_j
\end{bmatrix}$ have rank
\begin{equation}
	r_{ij}^* = 2n_1n_2-[2n_1-m_1]^+[2n_2-m_2]^+,
\end{equation}
where $[\cdot]^+$ denotes the positive part of a number.
\end{lemma}
\begin{IEEEproof}
Using standard matrix properties (e.g., \cite{cline1979rank}), we can write 
\begin{multline}
 r(\begin{bmatrix}
\mathbf{B}_i \otimes \mathbf{A}_i & \mathbf{B}_j \otimes \mathbf{A}_j
\end{bmatrix}) = r(\mathbf{B}_i \otimes \mathbf{A}_i) + r(\mathbf{B}_j \otimes \mathbf{A}_j) - \\ \dim[\mathcal{R}(\mathbf{B}_i \otimes \mathbf{A}_i) \bigcap \mathcal{R}(\mathbf{B}_j \otimes \mathbf{A}_j)].
\label{eq::model::5}
\end{multline}
Applying Lemma \ref{lemma::intersectionSpaces} from Appendix \ref{Appendix:DimInter}, we obtain
\begin{multline}
r(\begin{bmatrix}
\mathbf{B}_i \otimes \mathbf{A}_i & \mathbf{B}_j \otimes \mathbf{A}_j
\end{bmatrix})  = r(\mathbf{B}_i \otimes \mathbf{A}_i) + r(\mathbf{B}_j \otimes \mathbf{A}_j) -  \\ \dim[\mathcal{R}(\mathbf{A}_i ) \bigcap \mathcal{R}(\mathbf{A}_j )] \cdot \dim[\mathcal{R}(\mathbf{B}_i ) \bigcap \mathcal{R}(\mathbf{B}_j )].
\label{eq::CCB::5} 
\end{multline}
Almost every matrix has full rank, so $r(\mathbf{B}_i \otimes \mathbf{A}_i) = r(\mathbf{B}_j \otimes \mathbf{A}_j) = n_1n_2$ almost everywhere, so we can rewrite \eqref{eq::CCB::5} as
\begin{multline}
 r(\begin{bmatrix}
\mathbf{B}_i \otimes \mathbf{A}_i & \mathbf{B}_j \otimes \mathbf{A}_j
\end{bmatrix})  = 2n_1n_2 - \dim[\mathcal{R}(\mathbf{A}_i ) \bigcap \mathcal{R}(\mathbf{A}_j )] \\ \cdot \dim[\mathcal{R}(\mathbf{B}_i ) \bigcap \mathcal{R}(\mathbf{B}_j )].
\label{eq::CCB::6} 
\end{multline}
Next, we study the three possible cases for \eqref{eq::CCB::6}.\\
\textbf{Case 1}: $n_2<m_2<2n_2$ and $ n_1 \leq \frac{m_1}{2}$. Here,
\begin{align*}
%\mathcal{R}(\mathbf{A}_i ) \bigcap  \mathcal{R}(\mathbf{A}_j ) &= \{ 0 \}\\
\dim[\mathcal{R}(\mathbf{A}_i ) \bigcap  \mathcal{R}(\mathbf{A}_j )] &= 0  \\
\dim[\mathcal{R}(\mathbf{B}_i ) \bigcap  \mathcal{R}(\mathbf{B}_j )] &= (2n_2-m_2)\\
r(\begin{bmatrix}
\mathbf{B}_i \otimes \mathbf{A}_i & \mathbf{B}_j \otimes \mathbf{A}_j
\end{bmatrix})  &= 2n_1n_2 
\end{align*}
\textbf{Case 2}: $n_2\leq \frac{m_2}{2}$ and $ n_1<m_1<2n_1$. Here,
\begin{align*}
%\mathcal{R}(\mathbf{B}_i ) \bigcap  \mathcal{R}(\mathbf{B}_j ) &= \{ 0 \}\\
\dim[\mathcal{R}(\mathbf{B}_i ) \bigcap  \mathcal{R}(\mathbf{B}_j )] &= 0\\
\dim[\mathcal{R}(\mathbf{A}_i ) \bigcap  \mathcal{R}(\mathbf{A}_j )] &= (2n_1-m_1)\\
r(\begin{bmatrix}
\mathbf{B}_i \otimes \mathbf{A}_i & \mathbf{B}_j \otimes \mathbf{A}_j
\end{bmatrix})  &= 2n_1n_2 
\end{align*}
\textbf{Case 3}: $n_2<m_2<2n_2$ and $ n_1<m_1<2n_1$. Here,
\begin{align*}
\dim[\mathcal{R}(\mathbf{A}_i ) \bigcap  \mathcal{R}(\mathbf{A}_j )] &= (2n_1-m_1)\\
\dim[\mathcal{R}(\mathbf{B}_i ) \bigcap  \mathcal{R}(\mathbf{B}_j )] &= (2n_2-m_2)\\
r(\begin{bmatrix}
\mathbf{B}_i \otimes \mathbf{A}_i & \mathbf{B}_j \otimes \mathbf{A}_j
\end{bmatrix})  &= 2n_1n_2 - (2n_1-m_1)(2n_2-m_2),
\end{align*}
where the first and second equalities for each case hold almost everywhere, and the third equality for each case follows from Lemma \ref{lemma::intersectionSpaces}. Combining the three cases yields the claim.
\end{IEEEproof}
Applying Lemma \ref{lemma::FinalRank} to Theorem \ref{thm:diversity.order}, an exact expression for the diversity order follows immediately.
\begin{corollary}
	For almost every classification problem $\mathbf{a}$, the diversity order is
	\begin{equation}
		d(\mathbf{a}) = n_1n_2-[2n_1-m_1]^+[2n_2-m_2]^+.
	\end{equation}
\end{corollary}

\subsection{Diversity Order Gap}
% In the symmetric case: $m_1=m_2$ and $n_1=n_2$, the diversity order is the same as that predicted for general subspaces in \cite{nokleby2015discrimination}. 
% \textbf{Diversity order for K-S subspaces}:
% \begin{equation}
% d_{\text{K-S}} = n^2-([2n-m]^+)^2
% \end{equation}
% \textbf{Diversity order for Standard subsapce \cite{nokleby2015discrimination}}:
% \begin{equation}
% d_{\text{STD}} = n^2-[2n^2-m^2]^+
% \end{equation}
Diversity order characterize the slope of error probability, higher the diversity order faster the decay of misclassification probability. Since the Kronecker-structured subspaces comes from a restricted set of subsapces, the error performance of these subspaces can be worse. Therefore, to verify the efficiency of Kronecker subspaces, we characterizes the diversity order \emph{gap} as the difference between the slope of misclassification probability of K-S subspaces and the standard subspaces. This diversity order gap is a function of signal dimensions, that is, $n_1, n_2, m_1$ and $m_2$. We derive the signal dimension regimes where the diversity order gap is significant or/and zero.\\
\textbf{Diversity order for K-S subspaces}:
\begin{equation}
d_{\text{K-S}} = n_1n_2-[2n_1-m_1]^+[2n_2-m_2]^+.
\end{equation}
For the standard subspaces model in \eqref{eqn:class.conditional.density}, the signal of interest $\mathbf{Y} \in \mathbb{R}^{M}$ and coefficient vector $\mathbf{X} \in \mathbb{R}^{N}$ where, $M= m_1m_2$ and $N = n_1n_2$. From \cite{nokleby2015discrimination}, for the standard subspaces of same dimensions the diversity order would look like $N - [2N-M]^+$. This can be written in terms of Kronecker signal dimensions.\\ 
\textbf{Diversity order for standard subspace}:
\begin{equation}
d_{\text{STD}} = n_1n_2-[2n_1n_2-m_1m_2]^+.
\end{equation}
We observe that the diversity order for K-S models is never greater to the diversity order of standard subspace, for any value of $n_1, n_2, m_1,m_2$. However, for some regimes the diversity order of K-S model is smaller or equal to standard subspaces.\\
When $n_1<m_1<2n_1$ and  $n_2<m_2<2n_2$
\begin{enumerate}
\item \emph{if} $m_1m_2 > 2n_1n_2$ \emph{then} $d_{\text{K-S}} < d_{\text{STD}}$: $$\gamma= (2n_1-m_1)(2n_2-m_2).$$
\item \emph{if} $m_1m_2 < 2n_1n_2$ \emph{then} $d_{\text{K-S}} < d_{\text{STD}}$: $$\gamma= 2(m_1-n_1)(m_2-n_2),$$
\end{enumerate}
where $\gamma = d_{\text{STD}} - d_{\text{K-S}}$ is the diversity order gap. For any other region no diversity order gap exists, that is, $d_{\text{K-S}} = d_{\text{STD}}$. The details are provided in Appendix \ref{Appendix:gap}.

The high-SNR classification performance of K-S subspaces is the same as general subspaces when the subspace dimensions are small, even though K-S subspaces are structured, involve fewer parameters, and are easier to train. 
% With larger subspace dimensions, the K-S diversity order is smaller than that of unstructured subspaces.

\subsection{Misclassification Probability in terms of Row and Column Subspaces Geometry}
We derive a more accurate and tight high-SNR approximation of the probability of error in terms of principal angles between the K-S subspaces and also in terms of principal angle between the individual rows and columns subspaces. Using the eigenvalue decomposition of covariance of row subspace $\mathbf{A}_i\mathbf{A}_i^T = \mathbf{U}_i^A\lambda_i^A(\mathbf{U}_i^A)^T$ and the column subspace $\mathbf{B}_i\mathbf{B}_i^T = \mathbf{U}_i^B\lambda_i^B(\mathbf{U}_i^B)^T$, where $\mathbf{U}_i^A \in  \mathbb{R}^{m_1 \times n_1}, \mathbf{U}_i^B \in  \mathbb{R}^{m_2 \times n_2}$ are the orthonormal basis of row and column subspace respectively and the $\mathrm{diag}(\lambda_i^A) \in  \mathbb{R}^{n_1}, \mathrm{diag}(\lambda_i^B) \in  \mathbb{R}^{n_2}$ are the eigenvalues of row and column subspaces, we can write the signal covariance as:
\begin{align*}
\mathbf{D}_i\mathbf{D}_i^T &= ((\mathbf{U}_i^B\lambda_i^B(\mathbf{U}_i^B)^T \otimes \mathbf{U}_i^A\lambda_i^A(\mathbf{U}_i^A)^T))  \\
&= (\mathbf{U}_i^B\otimes \mathbf{U}_i^A ) (\lambda_i^B \otimes \lambda_i^A) ((\mathbf{U}_i^B)^T  \otimes (\mathbf{U}_i^A)^T) \\
&= (\mathbf{U}_i^B\otimes \mathbf{U}_i^A ) (\lambda_i^B \otimes \lambda_i^A) (\mathbf{U}_i^B \otimes \mathbf{U}_i^A)^T\\
&= \mathbf{U}_i\lambda_i\mathbf{U}_i^T 
\end{align*}
Similarly, $\mathbf{D}_j\mathbf{D}_j^T = \mathbf{U}_j\lambda_j\mathbf{U}_j^T$. From \cite{bernstein2009matrix}, the Kronecker product of two orthonormal matrix is a orthonormal matrix, thus $\mathbf{U}_i, \mathbf{U}_j \in \mathbb{R}^{m_1m_2 \times n_1n_2}$ are the orthonormal bases and the diagonal elements of $\lambda_i, \lambda_j \in \mathbb{R}^{n_1n_2 \times n_1n_2}$ are the eigenvalues. From equation \eqref{eq::model::5}, the rank of sum of two Kronecker products is written as:
% r^* &= r(\mathbf{D}_i\mathbf{D}_i^T + \mathbf{D}_j\mathbf{D}_j^T ) \nonumber \\
\begin{align}
 r^*& = r(\mathbf{D}_i\mathbf{D}_i^T) + r(\mathbf{D}_j\mathbf{D}_j^T) - \dim[\mathcal{R}(\mathbf{D}_i\mathbf{D}_i^T) \bigcap \mathcal{R}(\mathbf{D}_j\mathbf{D}_j^T)] \nonumber\\ 
 &= r(\mathbf{D}_i\mathbf{D}_i^T) + r(\mathbf{D}_j\mathbf{D}_j^T) - r_{\cap} \nonumber \\
&= 2n_1n_2 - r_{\cap}.\label{eq::Mis::1}
\end{align}
Since the intersection of two K-S subspaces define this rank and hence plays an important role in bounding the misclassification probability from above. According to \cite{huang2016role}, one can write the covariances of K-S subspaces in terms of subspaces intersections as follows:
\begin{align}
\Sigma_i =  \mathbf{U}_{i,\cap}\lambda_{i,\cap} \mathbf{U}_{i,\cap}^T + \mathbf{U}_{i,\setminus}\lambda_{i,\setminus} \mathbf{U}_{i,\setminus}^T + \sigma^2 \mathbf{I} ,\\
\Sigma_j = \mathbf{U}_{j,\cap}\lambda_{j,\cap} \mathbf{U}_{j,\cap}^T  + \mathbf{U}_{j,\setminus}\lambda_{j,\setminus} \mathbf{U}_{j,\setminus}^T + \sigma^2 \mathbf{I}.
\end{align}
Here $\mathbf{U}_{i,\cap}, \,\, \mathbf{U}_{j,\cap} \in \mathbf{R}^{m_1m_2, r_{\cap}}$ corresponds to the K-S subspace intersection and $\mathbf{U}_{i,\setminus}, \,\, \mathbf{U}_{j,\setminus} \in \mathbf{R}^{m_1m_2, n_1n_2 - r_{\cap}}$ corresponds to the set minus $\mathcal{D}_i \setminus \mathcal{D}_j$ and $\mathcal{D}_j \setminus \mathcal{D}_i$ respectively. Here $r_{\cap}$ accounts for the overlap between the subspaces, smaller the overlap between subspaces easier it to discern the classes. While on the other hand, $r_{\cap} = n_1n_2$ means the complete overlap between subspaces and it becomes hard to discriminate between classes. 
\begin{theorem}
	As $\sigma^2 \rightarrow 0$, the misclassification probability in terms of principal angle between individual row and column subspaces is upper bounded as 
	\begin{multline}
		P_e(\mathbf{D}_i,\mathbf{D}_j) \leq c_1 \left(\sigma^2\right)^{\frac{r^*_{ij}-n_1n_2}{2}} \\ \cdot\left(\prod_{l=t_1+1}^{n_1}\prod_{l=t_2+1}^{n_2}(1-\cos^2(\theta_l^A)\cos^2(\theta_l^B))\right)^{-\frac{1}{2}} \\+ o ((\sigma^2)^{\frac{r^*_{ij}-n_1n_2}{2}})
	\end{multline}
	where 
	\begin{multline}
	c_1 =  2^{\frac{n_1n_2-2}{2}} \cdot \left( \frac{\mathrm{pdet}(\mathbf{U}_{i,\cap}\lambda_{i,\cap} \mathbf{U}_{i,\cap}^T + \mathbf{U}_{j,\cap}\lambda_{j,\cap} \mathbf{U}_{j,\cap}^T)}{\sqrt{\prod_{l=1}^{r_{\cap}}\lambda_{i,\cap,l}\cdot \prod_{l=1}^{r_{\cap}}\lambda_{j,\cap,l} }} \right)^{-\frac{1}{2}}\\ \cdot \left(\sqrt{\prod_{l=1}^{n_1n_2-r_{\cap}}\lambda_{i,\setminus,l}\cdot \prod_{l=1}^{n_1n_2-r_{\cap}}\lambda_{j,\setminus,l}}\right)^{-\frac{1}{2}},
	\end{multline}
	$t_1 = \lfloor\frac{(n_2-n_1) - \sqrt{(n_2-n_1)^2 + 4r_{\cap}}}{2}\rfloor$,\\ $t_2 = \lfloor\frac{(n_1-n_2) - \sqrt{(n_1-n_2)^2 + 4r_{\cap}}}{2}\rfloor$ and $\mathrm{pdet}$ denotes the pseudo-determinant.
	\label{thm:Geometry}
\end{theorem}
\begin{IEEEproof}
Appendix \ref{Appendix:Geometry}.
\end{IEEEproof}
In case of no overlap between subspaces, that is, $r_{\cap} = 0$, both $t_1 = t_2 = 0$ and as the misclassification probability is inversely related to the product of all $n_1n_2$ principal angles, this makes the misclassification error negligibly small. On the other side, with subspace overlap $r_{\cap} \neq 0$, $t_1$ and $t_2$ has some positive value, there exists some non-trivial principal angles which effect the classification performance and it becomes very hard to distinguish between the subspaces.

\section{Classification capacity} \label{sect::Classification.capacity}

In this section, we derive upper and lower bounds on the classification capacity that hold approximately for large $\sigma^2$. Detailed analysis can be found in the long version of the paper.
\begin{theorem}
	The classification capacity is upper bounded by
	\begin{equation*}
		C \leq \frac{\min \{\nu_1,\nu_2\} (\kappa_1-\nu_1+\kappa_2-\nu_2)}{2\kappa_1\kappa_2}\log_2(1/\sigma^2) + O(1),
	\end{equation*}
	and
	\begin{equation*}
		C \geq \frac{\nu_1\nu_2-[2\nu_1-\kappa_1]^+[2\nu_2-\kappa_2]^+}{2\kappa_1\kappa_2}\log_2(1/\sigma^2) + O(1).
	\end{equation*}
\end{theorem}

\begin{IEEEproof}
The upper bound follows from an upper bound on the mutual information $I(\mathbf{y};\mathbf{A},\mathbf{B}) = h(\mathbf{y}) - h(\mathbf{y}|\mathbf{A},\mathbf{B})$ between the dictionary pairs $(\mathbf{A},\mathbf{B})$ and the signal $\mathbf{y}$
%Each class-conditional covariance is a rank-$n_1n_2$ matrix plus $\sigma^2\mathbf{I}$, so for small $\sigma^2$
%\begin{equation}
%	h(\mathbf{y}|\mathbf{A},\mathbf{B}) \approx \frac{m_1m_2 - n_1n_2}{2}\log_2(\sigma^2).
%\end{equation}
%By the i.i.d. Gaussian bound on differential entropy, $h(y) \leq m_1m_2/2\log_2(1+\sigma^2)$. Furthermore, when $n_1 < m_2$, the first $n_1$ rows of the matrix signal $\mathbf{Y}$ are enough to determine approxiately the column subspace basis $\mathbf{A}$, which leads to the estimate
%\begin{equation}
%	h(\mathbf{y}) \approx \frac{[m_2 - n_1]^+(m_1 - n_1)}{2}\log_2(\sigma^2).
%\end{equation}
%Similarly, when $n_2 < m_1$, the first $n_2$ columns of $\mathbf{Y}$ are sufficient to determine approximately the row subspace basis $\mathbf{B}$, which leads to the estimate
%\begin{equation}
%	h(\mathbf{y}) \approx \frac{[m_1 - n_2]^+(m_2 - n_2)}{2}\log_2(\sigma^2).
%\end{equation}
% Combining these cases and invoking Lemma \ref{lemma::classcapacity} gives the upper bound. The lower bound comes from taking a high-SNR approximation of the pairwise error as derived in Section \ref{sect::Diversity.order} and applying the union bound.
 and invoke Lemma \ref{lemma::classcapacity}. In particular,
 \begin{equation}
 	I(\mathbf{y};\mathbf{A},\mathbf{B}) = h(\mathbf{y}) - h(\mathbf{y}|\mathbf{A},\mathbf{B}).
 \end{equation}
 Given the conditional distribution $p(\mathbf{y}|(\mathbf{B}\otimes \mathbf{A})) = \mathcal{N}(0, (\mathbf{B}\otimes \mathbf{A})(\mathbf{B}\otimes \mathbf{A})^T + \sigma^2 \cdot \mathbf{I})$ and following the analysis similar to \cite{nokleby2015discrimination} we bound the conditional entropy as:
 % Let $\lambda_i$ be the $i$th eigenvalue of $(\mathbf{B}\otimes \mathbf{A})(\mathbf{B}\otimes \mathbf{A})^T$; then the conditional entropy is
 % \begin{multline}
 % h(\mathbf{y}|\mathbf{A},\mathbf{B}) = \sum_{i=1}^{n_1n_2}\frac{1}{2}E[\log_2(2\pi e(\lambda_i+\sigma^2))] \\ + \frac{m_1m_2 - n_1n_2}{2}\log_2(2\pi e\sigma^2),
 % \end{multline}
 % which is bounded by
 \begin{multline}
 h(\mathbf{y}|\mathbf{A},\mathbf{B}) \geq \frac{m_1m_2 - n_1n_2}{2}\log_2(\sigma^2)+\frac{m_1m_2}{2}\log_2(2\pi e)\\ + \frac{n_1n_2}{2}E[\log_2((\sqrt{\kappa_1/\upsilon_1} - 1)^2 \cdot (\sqrt{\kappa_2/\upsilon_2} - 1)^2 + \epsilon(m)+ \sigma^2)] ,
 \label{eq::CCB::7}
 \end{multline}
% %\begin{equation}
% %h(Y|A,B) \approx \frac{m_1m_2 - n_1n_2}{2}\log(\sigma^2)
% %\end{equation}
% %with high probability.\\
 From the i.i.d. Gaussian outer bound on entropy, we can derive a naive bound on the marginal entropy:
 \begin{equation}
 	h(\mathbf{y})  \leq  \frac{m_1m_2}{2}\log(1+\sigma^2)
 \end{equation}
 Now consider the case when both $\mathbf{A}$ and $\mathbf{B}$ are tall i.e. $ m_1 >n_1 $ and $ m_2 > n_2$. Further suppose that $n_1 < m_2$. Then, we can derive a tighter outer bound on $h(\mathbf{y})$. 
 Let $\mathbf{y}_p$ be the first $n_1$ columns of $\mathbf{y}$ and let $\mathbf{y}^\prime_p$ be the rest $m_2-n_1$ columns of $\mathbf{y}$. Then, $\mathbf{y}^\prime_p \in \mathbb{R}^{m_1 \times (m_2-n_1)}$, and we can derive the following high-SNR approximation on $h(\mathbf{y})$:
 \begin{align}
 h(\mathbf{y}_\mathbf{A})  &=  h(\mathbf{y}_p)+h(\mathbf{y}^\prime_p|\mathbf{y}_p)\\
  & \simeq  h(\mathbf{y}_p)+h(\mathbf{y}^\prime_p|\mathbf{A})
 \end{align}
 \begin{equation}
 \label{eq::CCB::2}
 h(\mathbf{y}_\mathbf{A}) = \frac{m_1n_1}{2}\log_2(1+\sigma^2) + \frac{[m_2 - n_1]^+(m_1 - n_1)}{2}\log_2(\sigma^2)
 \end{equation}
 Now, let $\mathbf{y}_q$ be the first $n_2$ columns of $\mathbf{y}$ and let $\mathbf{y}^\prime_q$ denotes the rest $m_1-n_2$ columns of $\mathbf{y}$. Then, $\mathbf{y}^\prime_p \in \mathbb{R}^{(m_2 - n_2) \times m_2}$, and we derive the following high-SNR approximation on $h(\mathbf{y})$:
\begin{align}
 % h(\mathbf{y}_\mathbf{B})  &=  h(\mathbf{y}_q)+h(\mathbf{y}^\prime_q|\mathbf{y}_q)\\
 h(\mathbf{y}_\mathbf{B}) & \simeq  h(\mathbf{y}_q)+h(\mathbf{y}^\prime_q|\mathbf{B}) 
 \end{align}
 \begin{equation}
  \label{eq::CCB::3}
  h(\mathbf{y}_\mathbf{B})  = \frac{m_2n_2}{2}\log_2(1+\sigma^2) + \frac{[m_1 - n_2]^+(m_2 - n_2)}{2}\log_2(\sigma^2)
 \end{equation}
 Combining \eqref{eq::CCB::2} and \eqref{eq::CCB::3}, we obtain the differential entropy:
 \begin{align}
 h(\mathbf{y})  &\leq  \min (h(\mathbf{y}_\mathbf{A}),h(\mathbf{y}_\mathbf{B}))\\
&=  \frac{\min \{(m_2 - n_1)(m_1 - n_1),(m_1 - n_2)(m_2 - n_2)\}}{2} \nonumber \\ & \quad \times \log_2(\sigma^2) + \frac{\min \{m_1n_1,m_2n_2\}}{2}  \log_2(1+\sigma^2) 
 \label{eq::CCB::8}
 \end{align}
 From \eqref{eq::CCB::7} and \eqref{eq::CCB::8}, as $m \rightarrow \infty$ we can find the bound. 
\end{IEEEproof} 
\begin{IEEEproof}[\textbf{Lower Bound}]
 In order to obtain the lower bound on classification capacity we apply the Bhatacharyya bound on probability of pairwise error between two Kronecker-subspaces $i$ and $j$. By expanding $r^*_{ij}$ in \eqref{eq::CCB::11}
 % and obtained $P_e(\mathbf{a})$ by expandingfrom \eqref{eq::CCB::11} as follows:
 % %\begin{figure*}[!h]
 % \begin{multline}
 % \log_2(E[P_e(\mathbf{a})]) \leq \rho m_1m_2 \\+\frac{n_1n_2-2}{2} -\frac{n_1n_2-[2n_1-m_1]^+[2n_2-m_2]^+}{2} \\ \cdot \log_2\left( 1+\frac{\lambda_{ij\{2n_1n_2-[2n_1-m_1]^+[2n_2-m_2]^+\}}}{\sigma^2}\right)
 % \end{multline}
 %\end{figure*}
 and bounding the value of $\lambda_{ij\{2n_1n_2-[2n_1-m_1]^+[2n_2-m_2]^+\}}$  away from zero as $m\rightarrow \infty$. If
 \begin{multline}
 \rho < \frac{n_1n_2-2}{2m_1m_2}-\frac{n_1n_2-[2n_1-m_1]^+[2n_2-m_2]^+}{2m_1m_2} \times \\ \log_2\left( 1+\frac{\lambda_{ij\{2n_1n_2-[2n_1-m_1]^+[2n_2-m_2]^+\}}}{\sigma^2}\right),
 \end{multline}
 then surely $P_e(\mathbf{a})$ goes to zero as $m \rightarrow \infty $.
 %In high signal to noise setting, that is $\sigma^2 \rightarrow 0$
 %\begin{align}
 %P_e(\mathbf{D}_i,\mathbf{D}_j) &\leq \left(\frac{1}{\sigma^2}\right)^{-\frac{min \{m_1m_2 - 2(m_1-n_1)(m_2-n_2),2n_1n_2\}-n_1n_2}{2}} + C \\
 %&\leq \left(\frac{1}{\sigma^2}\right)^{-\frac{min \{m_1m_2-n_1n_2 - 2(m_1-n_1)(m_2-n_2),n_1n_2\}}{2}} +C 
 %\end{align}
 %Here, we find the pre-log term as $$\frac{min \{m_1m_2-n_1n_2 - 2(m_1-n_1)(m_2-n_2),n_1n_2\}}{2} $$
\end{IEEEproof}

To compare the upper and lower bounds, consider the symmetric case, i.e. $m_1 = m_2 = m$ and $n_1=n_2=n$ and $m>n$. The gap between the prelog factor of the upper and lower bounds is $(m-n)^2$ and we leave tightening these bounds as future work.
% We believe that both the upper and lower bounds are loose in general, and future work will involve tighter bounds on the mutual information in order to improve the prelog estimates.

\section{Kronecker-Structured Learning of Discriminative Dictionaries (\mbox{K-SLD$^2$})} \label{sect::Dictionary.Learning}

Here we introduce K-SLD$^2$, an efficient and effective method for learning discriminative dictionary pairs for classifying two-dimensional signals $~{\mathbf{Y} \in \mathbf{R}^{m_1 \times m_2}}$ in \eqref{eq::model}.
%  In particular, we suppose that the signals approximately have the following form:
%  \begin{equation}\label{eqn:2d.model}
% \mathbf{Y} = \mathbf{A}\mathbf{X}\mathbf{B}^T,
% \end{equation}
% where $\mathbf{X} \in \mathbb{R}^{n_1\times n_2}$ is the coefficient matrix and $\mathbf{A}\in \mathbb{R}^{m_1\times n_1}$  describes the subspace on which columns of $\mathbf{Y}$ approximately lies and $\mathbf{B}\in \mathbf{R}^{m_2\times n_2}$ describes the subspace on which rows of $\mathbf{Y}$ approximately lies \cite{shakeri2016minimax}. %So, $Y$ can be written as $Y = AXB^T$. 
%In supervised learning, a common approach to dictionary learning is to learn class-specific dictionaries for each class. To classify a test signal, simply choose the class whose dictionary best reconstructs the test signal.
For $L$ number of classes let $K$ is the number of training samples per class. We define $Y_i$ as a collection of $K$ 2-D signals corresponding to class $i$. That is, $$Y_i = \{\mathbf{Y}_{1i}, \mathbf{Y}_{2i}, \cdots , \mathbf{Y}_{Ki}\},$$ for $i=1,\cdots,L$ and $\mathbf{Y}_{ji} \in \mathbf{R}^{m_1 \times m_2}$ is the $j$th signal belonging to class $i$. 

We suppose that each class corresponds to a different subspace. Thus, our objective is to learn the structured dictionary pairs $\mathbf{\mathbb{A}} = \{\mathbf{A}_1,\mathbf{A}_2,\cdots, \mathbf{A}_L\}$ and $\mathbf{\mathbb{B}}  = \{\mathbf{B}_1,\mathbf{B}_2,\cdots, \mathbf{B}_L\}$ that describe the training data. We define the set of structured dictionary pairs as $(\mathbf{\mathbb{A}}, \mathbf{\mathbb{B}}) = \{(\mathbf{A}_1, \mathbf{B}_1), (\mathbf{A}_2, \mathbf{B}_2), \cdots, (\mathbf{A}_L, \mathbf{B}_L)\}$, where $(\mathbf{A}_i, \mathbf{B}_i)$ is the class-specific sub-dictionary pair associated with class $i$. 

% The term {\em Kronecker-structured} refers to the form of the dictionaries that results from vectorizing the signal model (\ref{eqn:2d.model}):
% \begin{equation}\label{eqn:kronecker.structure}
% 	\mathbf{y} = (\mathbf{B}\otimes \mathbf{A})\mathbf{x},
% \end{equation}
% where $\mathbf{y} = \mathrm{vec}(\mathbf{Y})$, $\mathbf{x} = \mathrm{vec}(\mathbf{X})$ and $\otimes$ represents the Kronecker product. Analogous to standard subspace learning, we can consider the signal to live approximately on a union of subspaces defined by the dictionary $\mathbf{D}_i = (\mathbf{B}_i\otimes \mathbf{A}_i)$.

Let $\mathbf{\mathbb{X}} = \{S_1,S_2,\cdots, S_L\}$ be a set of coefficient matrices for each signal, where $S_i = \{X_{1i}, X_{2i}, \dots , X_{Ki}\}$ is the sub-matrix containing the coefficients of all the training samples $Y_i$ belongs to a class $i$ over the dictionary pair $(\mathbf{\mathbb{A}}, \mathbf{\mathbb{B}})$. We write, $X_{ji} = \{\mathbf{X}_{ji}^1, \mathbf{X}_{ji}^2, \cdots, \mathbf{X}_{ji}^L \}$ a representation of signal $j$ of class $i$ over the dictionary pair $(\mathbf{\mathbb{A}}, \mathbf{\mathbb{B}})$, where $\mathbf{X}_{ji}^l \in \mathbb{R}^{n_1\times n_2}$ is the coefficient of a training sample $\mathbf{Y}_{ji}$ over the dictionary pair $(\mathbf{A}_k, \mathbf{B}_k)$. That is, $(\mathbf{\mathbb{A}}, \mathbf{\mathbb{B}})$ represent an overcomplete dictionary, and we learn coefficients such that
\begin{equation}
	\mathbf{Y}_{ji} = \sum_{l=1}^L\sum_{j=1}^K \mathbf{A}_i \mathbf{X}_{ji}^l\mathbf{B}_i^T.
\end{equation} 

 %We denote the representation of a training sample $\mathbf{Y}_{ji}$ over the dictionary pair $(\mathbf{A}_k, \mathbf{B}_k)$ as $\mathbf{R}_{ji}^k = \mathbf{A}_k\mathbf{X}_{ji}^k\mathbf{B}_k^T$. Then, the set of structured dictionary pair $(\mathbf{\mathbb{A}}, \mathbf{\mathbb{B}})$ describes the training data $\mathbf{Y}_{ji}$ well if there exists a coefficient matrix $\mathbf{X}_{ji}^i$ such that ~$||\mathbf{Y}_{ji} - \sum_{k=1}^K\mathbf{R}_{ji}^k||_F^2$ is small.

\textbf{Algorithm Description:} We want the dictionaries to have both high reconstruction power and high {\em discriminative} power. To encourage discriminability, we want a signal $\mathbf{Y}_i$ to be well represented by the class-specific dictionary $(\mathbf{A}_i,\mathbf{B}_i)$, and (comparatively) poorly represented by the other dictionaries $(\mathbf{A}_l,\mathbf{B}_l), l\neq i$. Here, $\mathbf{A}_l\mathbf{X}_{ji}^l\mathbf{B}_l^T$ denotes the representation of the training sample $\mathbf{Y}_{ji}$ over the $l$th dictionary pair. Then, the dictionaries discriminate well if $||\mathbf{Y}_{ji} - \mathbf{A}_l\mathbf{X}_{ji}^l\mathbf{B}_l^T||_2$ is small for $i=l$ and large for $i \neq l$. This leads to a optimization problem:
\begin{multline}
\min_{\{\mathbb{A}, \mathbb{B}, \mathbb{X}\}} \sum_{i=1}^L\sum_{j=1}^K \Bigg( ||\mathbf{Y}_{ji} - \sum_{l=1}^L\mathbf{A}_l\mathbf{X}_{ji}^l\mathbf{B}_l^T||_F^2 + \\||\mathbf{Y}_{ji} - \mathbf{A}_i\mathbf{X}_{ji}^i\mathbf{B}_i^T||_F^2  + \mu \sum_{l=1, l\neq i}^L ||\mathbf{A}_l\mathbf{X}_{ji}^l\mathbf{B}_l^T||_F^2 \Bigg).
\label{eq::Opt}
\end{multline}
The first term in (\ref{eq::Opt}) encourages the representation power of the joint, overcomplete dictionary, whereas the second and third terms encourage the discrimination power of the class-specific dictionaries. 
This problem is jointly nonconvex, but it is convex in the individual variables $\mathbb{A},\mathbb{B},\mathbb{X}$ when the other are fixed. We solve \eqref{eq::Opt} by alternating between the variables, solving the individual convex problem, and iterating until convergence. Thus, we divide \eqref{eq::Opt} into three subproblems: updating $\mathbb{X}$ while fixing $\mathbb{A}$ and $\mathbb{B}$; updating $\mathbb{A}$ while fixing $\mathbb{X}$ and $\mathbb{B}$; and updating $\mathbb{B}$ while fixing $\mathbb{X}$ and $\mathbb{A}$. Each subproblem further has a closed-form solution. The solution to the first subproblem is
% \begin{equation*}
% \mathbf{A}_i^* = \frac{\sum_{j=1}^{n_i}\left(4\mathbf{Y}_{ji} - \sum_{k=1, k\neq i}^K \mathbf{A}_k\mathbf{X}_{ji}^k\mathbf{B}_k^T\right)\mathbf{B}_i\mathbf{X}_{ji}^i }{4 \sum_{j=1}^{n_i} \mathbf{X}_{ji}^i\mathbf{B}_i^T(\mathbf{X}_{ji}^i\mathbf{B}_i^T)^T}.
% \end{equation*}
\begin{multline}
\mathbf{A}_i^* = \frac{1}{4}\sum_{j=1}^{K}\left(4\mathbf{Y}_{ji} - \sum_{l=1, l\neq i}^L \mathbf{A}_l\mathbf{X}_{ji}^l\mathbf{B}_l^T\right)\mathbf{B}_i\mathbf{X}_{ji}^i \times \\ \left(\sum_{j=1}^{N} \mathbf{X}_{ji}^i\mathbf{B}_i^T(\mathbf{X}_{ji}^i\mathbf{B}_i^T)^T\right)^{-1}.
\label{eq::updateA}
\end{multline}
Then, the solution to the second subproblem is
% \begin{equation*}
% \mathbf{B}_i^* = \frac{\sum_{j=1}^{N}\left(4\mathbf{Y}_{ji}^T - \sum_{k=1, k\neq i}^K \mathbf{B}_k(\mathbf{A}_k^*\mathbf{X}_{ji}^k)^T\right)\mathbf{A}_i^*\mathbf{X}_{ji}^i }{4 \sum_{j=1}^{N} (\mathbf{A}_i^*\mathbf{X}_{ji}^i)^T\mathbf{A}_i^*\mathbf{X}_{ji}^i}.
% \end{equation*}
\begin{multline}
\mathbf{B}_i^* = \frac{1}{4} \sum_{j=1}^{K}\left(4\mathbf{Y}_{ji}^T - \sum_{l=1, l\neq i}^L \mathbf{B}_l(\mathbf{A}_l^*\mathbf{X}_{ji}^l)^T\right)\mathbf{A}_i^*\mathbf{X}_{ji}^i \times \\  \left(\sum_{j=1}^{N} (\mathbf{A}_i^*\mathbf{X}_{ji}^i)^T\mathbf{A}_i^*\mathbf{X}_{ji}^i\right)^{-1}.
\label{eq::updateB}
\end{multline}
Finally, the solution to the third subproblem is, for $i  = l$
\begin{multline}
(\mathbf{X}_{ji}^i)^* = \frac{1}{2}\left((\mathbf{A}_i^*)^T\mathbf{A}_i^*\right)^{-1}(\mathbf{A}_i^*)^T \times \\ \sum_{j=1}^{K}\left(4\mathbf{Y}_{ji} - \sum_{l=1, l\neq i}^L \mathbf{A}_l^*\mathbf{X}_{ji}^l(\mathbf{B}_l^*)^T\right) \mathbf{B}_i^*\left((\mathbf{B}_i^*)^T\mathbf{B}_i^*\right)^{-1},
\label{eq::updateXi}
\end{multline}
and for $i \neq l$:
\begin{multline}
(\mathbf{X}_{ji}^l)^* = \frac{1}{2}\left((\mathbf{A}_l^*)^T\mathbf{A}_l^*\right)^{-1}(\mathbf{A}_l^*)^T \times \\\sum_{j=1}^{K}\left(2\mathbf{Y}_{ji} - \sum_{t=1, t\neq l}^L \mathbf{A}_t^*\mathbf{X}_{ji}^t(\mathbf{B}_t^*)^T\right) \mathbf{B}_l^*\left((\mathbf{B}_l^*)^T\mathbf{B}_l^*\right)^{-1}.
\label{eq::updateXj}
\end{multline}
These iterations continue until changes in the objective function are sufficiently small.

\begin{figure}[!h]
\centering
\begin{subfigure}[b]{0.49\columnwidth}
  \centering
  \includegraphics[width=\textwidth]{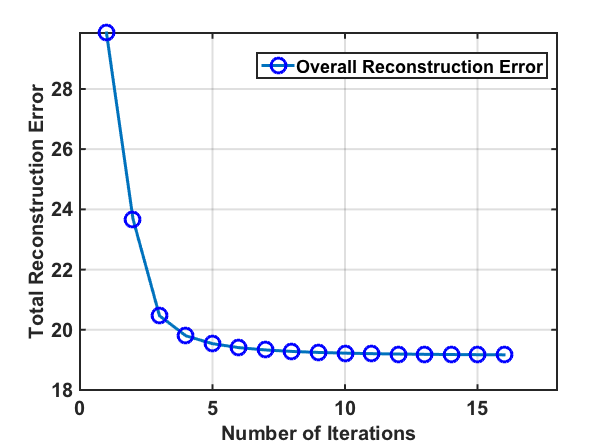}
  \caption{}
  \label{fig::TotalR}
\end{subfigure}
\begin{subfigure}[b]{0.49\columnwidth}
  \centering
  \includegraphics[width=\textwidth]{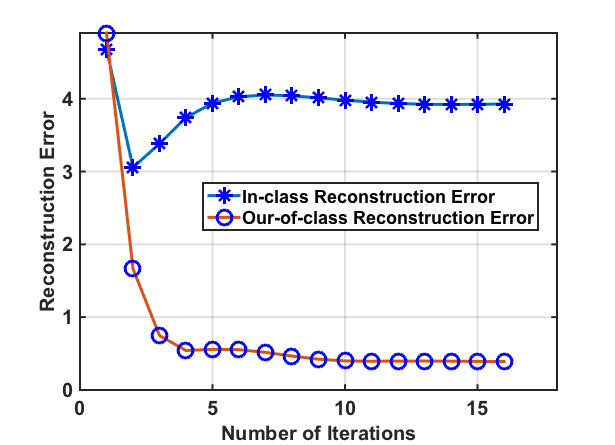}
  \caption{}
  \label{fig::Reconstruction}
\end{subfigure}
\caption{Convergence performance on extended YaleB face recognition dataset. (a) shows the overall reconstruction error; (b) shows the in-class and the out-of-class reconstruction error.}
\label{fig::Convergence}
\end{figure}

\textbf{Convergence:} This procedure is guaranteed to converge in terms of the objective function value via the following argument. Because each subproblem is convex, the value of the objective function is nondecreasing as iterations proceed. Furthermore, because the objective function is bounded below, the nondecreasing sequence of function values must converge. A sample trajectory is shown in Fig. \ref{fig::TotalR}. Here, K-SLD$^2$ is trained on the extended YaleB dataset. The overall reconstruction error is shown in \eqref{eq::Opt},  whereas Fig. \ref{fig::Reconstruction}, shows both that the signal $Y_i$ is well represented by the dictionary pair $(\mathbf{A}_i,\mathbf{B}_i)$ and as the number of iterations increases the other dictionary pairs $(\mathbf{A}_l,\mathbf{B}_l), l\neq i$ start loosing their ability to represent $Y_i$. 

\textbf{Classification Procedure:} Given a test signal $\mathbf{Y}$ to classify, we first find the coefficient matrices for each class using
\begin{equation}
 \{\mathbf{\hat{X}}^i\} = \arg\,\min_{\{\mathbf{X}^i\}_{i=1}^L} ||\mathbf{Y} - \sum_{l=1}^L\mathbf{A}_l\mathbf{X}^l\mathbf{B}_l^T||_F^2.
 \label{eq::TestX}
\end{equation}
This problem is convex and has a closed-form solution. Then, we compute the reconstruction error for each class-specific dictionary:
\begin{equation}
 e_i = ||\mathbf{Y} - \mathbf{A}_i\mathbf{\hat{X}}^i\mathbf{B}_i^T||_F^2.
 \label{eq::TestReconstruction}
\end{equation}
Finally, we make the prediction $\hat{k} = \arg\min_{i=1,\cdots,L} (e_i)$; i.e., the class with the smallest reconstruction error.

\textbf{Computational Complexity:} In this analysis we use the fact that: 1) if $\mathbf{A}\in \mathbb{R}^{m_1\times n_1}$ and $\mathbf{X}\in \mathbf{R}^{n_1\times n_2}$ then the matrix multiplication $\mathbf{A}\mathbf{X}$ has complexity $m_1n_1n_2$. 2) if a non singular matrix $\mathbf{A} \in \mathbb{R}^{n_1 \times n_1}$, then $\mathbf{A}^{-1}$ has complexity $n_1^3$. We obtain a complexity (in terms of matrix multiplications and additions) of $$\mathcal{O}(KLn_1m_2(m_1+n_2)).$$ If we assume $m_1 = m_2 = \sqrt{m}$ and $n_1 = n_2 = \sqrt{n}$ then the complexity becomes $$\mathcal{O}(KL(m\sqrt{n}+n\sqrt{m})).$$ Which is a reduction when compared to standard subspace learning with computational complexity of $\mathcal{O}(KLnm).$

\section{Numerical Results} \label{sect::Numerical.Result}

In this section, we evaluate first demonstrate that the empirical classification performance, when the classes are perfectly known, agrees with the diversity order and bounds derived above. Then, we demonstrate the learning and classification performance of \mbox{K-SLD$^2$} on both synthetic and real-world data.

\subsection{Diversity Order}
\subsubsection{Synthetic Data}
\begin{figure}[h]
  \centering
    \includegraphics[width=0.4\textwidth]{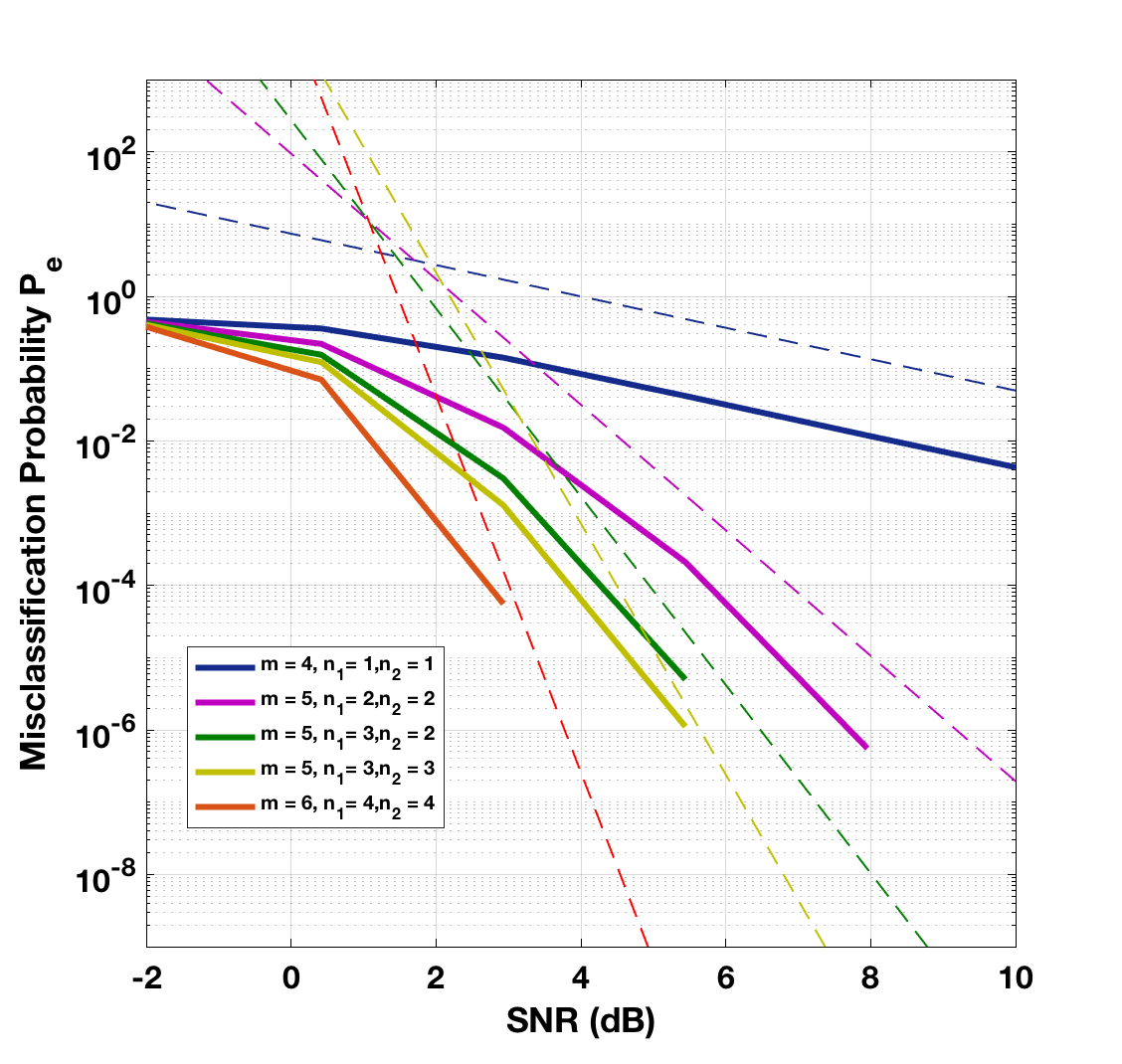}
   \caption{Misclassification probability $P_e$ Vs. SNR}
    \label{fig::Numerical::syn}
\end{figure}

We randomly choose two classes by drawing matrix pairs $\mathbf{A}_i$ and $\mathbf{B}_i$ independently from the distribution in (\ref{eqn:model.prior}). Then, we draw data samples i.i.d. from the class-conditional densities in (\ref{eqn:class.conditional.density}). We classify each data sample by minimizing the Mahalanobis distance  associated with the covariance of each class-conditional density. We consider five cases, in which we fix $m_1=m_2=m$ and vary $n_1$ and $n_2$. In Fig.\ref{fig::Numerical::syn} we plot the misclassification probability $P_e$ against the SNR in dB, averaged over $10^5$
%[ISHAN: how many?] 
random draws from each class. We also plot the upper bound on misclassification probability in terms of principal angles for each case described in \eqref{Eq:principal}. Where dotted colored line shows the misclassification probability associated with the corresponding solid line for each case. In each case, the empirical performance agrees with the diversity predictions with an offset. This offset is large when the ambient signal dimension is small and with large dimensions this offset approaches to zero.
% slope predicted by the diversity order for each case, which is as small as 1 and as large as 12. In each case, the empirical performance agrees with the diversity predictions. For larger values of $m$, the diversity order is sufficiently high that it is difficult to estimate $P_e$ reliably.

%Diversity order is 12. For the higher values of $m$ we observed that the slope of the curve for small values of noise tends to negative infinity.

 % \subsubsection{YaleB Faces}
 % Now, we present the performance of our analysis on YaleB face dataset. We randomly choose 10 classes out of the 38 face classes in the set. For each class we learn the Kronecker-structured dictionary that best fits the dataset images. Then, we project the images from each class onto its learned Kronecker subspace. This enforces the Kronecker structure on the images, which makes it possible to evaluate the diversity performance. We calculate the misclassification probability via minimizing the residual error between the image and its noisy Kronecker subspace projection
 % %[ISHAN: What is the residual method? This is vague.] 
 % for 10,000 noisy instantiations of each image. In this problem, the ambient dimensions $m = 32$. We show results for $n_1 = 4$ and $n_2 = 2$, which leads to a diversity order sufficiently small that we can estimate $P_e$ reliably. For this particular case when $m = 32, n_1 = 4$ and $n_2 = 2$ diversity order is 8 and the misclassification probability is plotted with respect to SNR in Figure \ref{fig::Numerical::real}. The empirical performance is similar to theoretical predictions.
 % \begin{figure}[h]
 %   \centering
 %     \includegraphics[width=0.45\textwidth]{Yale_b_4_2_D_ed.png}
 %    \caption{Misclassification probability $P_e$ Vs. SNR}
 %     \label{fig::Numerical::real}
 % \end{figure}

\subsection{Dictionary Learning Algorithm}
 In this section we evaluate the performance of \mbox{K-SLD$^2$} algorithm on synthetic data and two real world datasets: extended YaleB face dataset \cite{georghiades2001few} and the UCI EEG dataset \cite{zhang1995event}, which differentiates the EEG signals of control patients and those who suffer from alcoholism. We compare the performance to state-of-the-art dictionary learning methods such as FDDL \cite{yang2011fisher}, DLSI \cite{ramirez2010classification}, LRSDL \cite{vu2017fast}, standard subspace learning (SSL) using \eqref{eq::vectPre} as a baseline method, and the standard kernel support vector machine (SVM). We perform learning and classification on unprocessed signals. When appropriate, we choose model hyper-parameters via cross-validation.

  %and a popular classification method kernel \emph{support vector machine (SVM)}. In all of our experiments we did not apply any transformations to the input signal such as in \cite{ramirez2010classification,yang2011fisher} the dimension of input faces image is usually reduced to 504 using random-face features \cite{turk1991eigenfaces}. For the fair comparison, we input the raw face images as an input to all the methods without any feature engineering. For the methods \cite{ramirez2010classification,yang2011fisher,vu2017fast}, we choose the hyper-parameters associated with the best performance.

\subsubsection{Synthetic data}
\begin{figure}[h]
  \centering
    \includegraphics[width=0.35\textwidth]{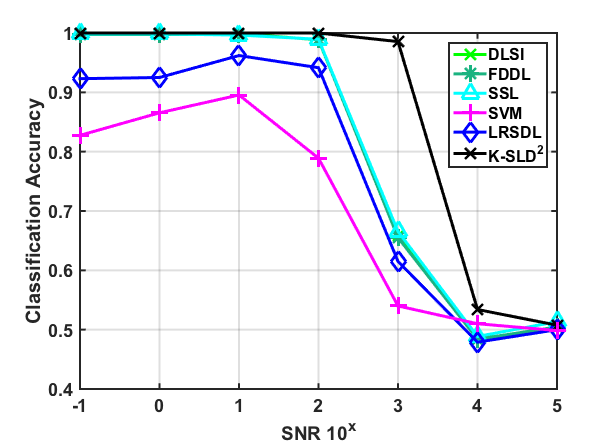}
   \caption{Classification accuracy Vs. SNR for synthetic data}
    \label{fig::synthetic}
\end{figure}

We consider two class classification problem where we draw two matrix pairs $\mathbf{A}_i$ and $\mathbf{B}_i$ independently from \eqref{eqn:model.prior} and draw data samples i.i.d from the class-conditional densities in \eqref{eqn:class.conditional.density}. For this experiment we choose the dimensions of the signal to be $32 \times 32$ which lies on the row and column subspaces of dimension $13$ and $17$, respectively. For each class we draw 10 samples for training/dictionary learning and 50 samples for testing. In total we have 60 samples per class. For learning K-S dictionaries using \mbox{K-SLD$^2$} we use the 2-D signal as it is while for the other learning algorithms we first vectorize the signal (dimension $1024\times 1$). Fig. \ref{fig::synthetic} compares the performance of learned dictionaries using different methods as the SNR decreases. When the noise power is low, that is, $\leq 10^1$, standard subspace learning and \mbox{K-SLD$^2$} performs equally well, but as the noise power increases a significant gain in performance is observed as evident in Fig. \ref{fig::synthetic}. We find best classification performance for SVM with polynomial kernel of degree 3. 

\subsubsection{Face Recognition}
\begin{figure}[htbp]
\centering
\begin{subfigure}[b]{0.45\columnwidth}
  \centering
  \includegraphics[width=\textwidth]{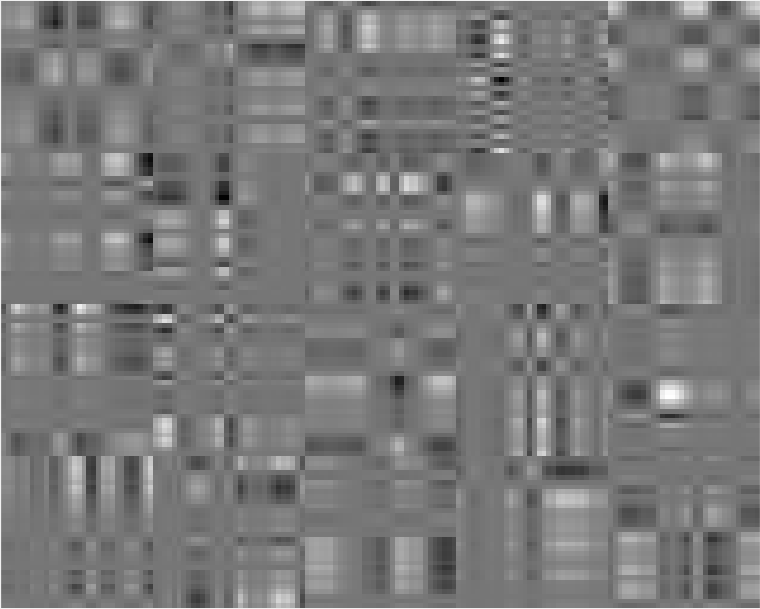}
  \caption{}
  \label{fig::KSDict}
\end{subfigure}
\begin{subfigure}[b]{0.45\columnwidth}
  \centering
  \includegraphics[width=\textwidth]{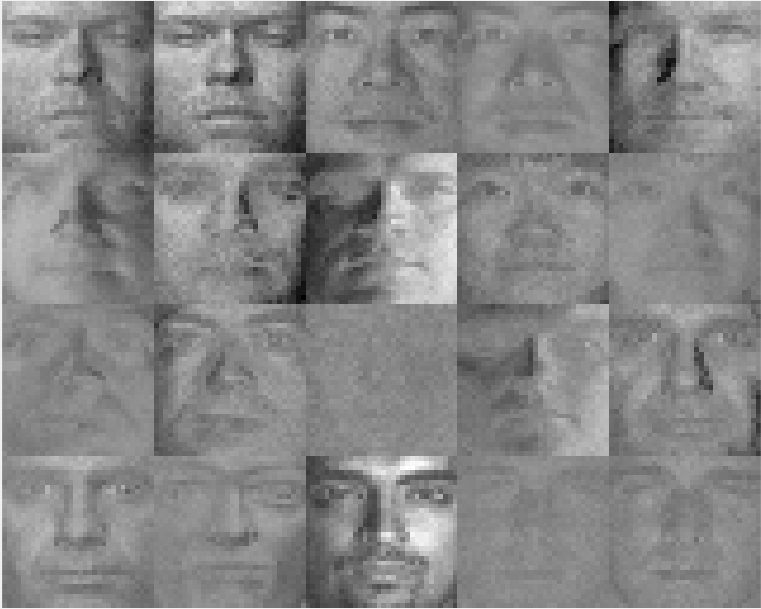}
  \caption{}
  \label{fig::StdDict}
\end{subfigure}
\caption{A subset of dictionary atoms learned by (a) \mbox{K-SLD$^2$} model, and  (b) standard subspace model. }
\label{fig::Dict}
\end{figure}

The extended YaleB dataset consists of 2,414 frontal face images from 38 individuals captured under varying lighting conditions. %We learn K-S dictionary pair $(\mathbf{A},\mathbf{B})$ for each class. Where a signal is represented by sparse matrix coefficients in a dictionary pair. A dictionary pair describes the training data well if there exists a coefficient matrix which minimizes the reconstruction error.  
%[As discussed, make this more explicit. Describe in words the components of the objective function.] For the signals from different class to live in K-S subspace that are far apart we add an incoherence term between dictionaries. 
For each class, we use 10 images for training/dictionary learning and the remaining 54 images for testing. In Figure \ref{fig::Dict} we show the dictionaries learned by \mbox{K-SLD$^2$} vs. a standard subspace learning model, and we observe that the standard model learns dictionary atoms that look similar to a few reference faces for each class, whereas the \mbox{K-SLD$^2$} learns more abstract dictionary atoms. This is in part due to imposition of the Kronecker structure on the dictionary atoms, as well as the larger number of atoms possible in a K-S dictionary.
\begin{figure}[htbp]
\centering
\begin{subfigure}[b]{0.35\textwidth}
  \centering
  \includegraphics[width=\textwidth]{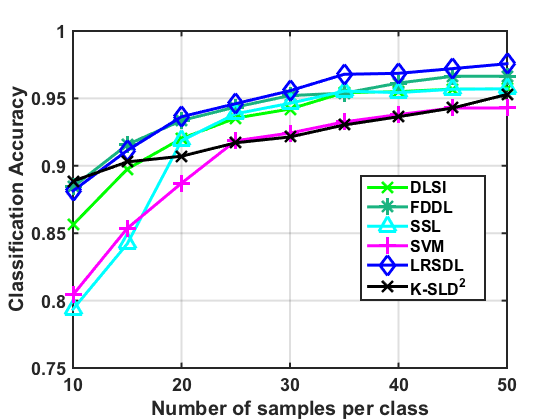}
  \caption{}
  \label{fig::YaleAccuracy}
\end{subfigure}
\begin{subfigure}[b]{0.35\textwidth}
  \centering
  \includegraphics[width=\textwidth]{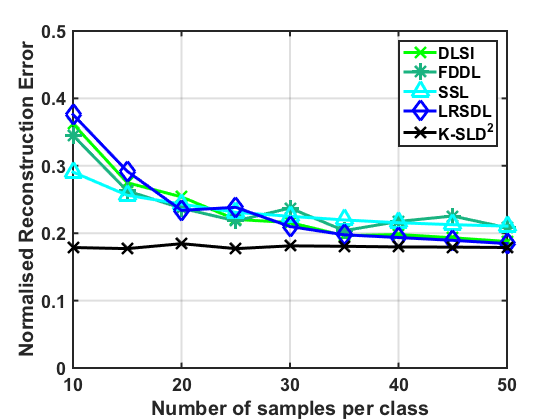}
  \caption{}
  \label{fig::YaleRecon}
\end{subfigure}
\caption{Performance on extended YaleB dataset (a) classification accuracy (b) normalized reconstruction error}
\label{fig::Yale_Performance}
\end{figure}
\begin{table*}[ht]
\centering
\begin{tabular}{|c|c|c|c|c|c|c|}
\hline
  & \begin{tabular}[c]{@{}c@{}}SSL\end{tabular} & \begin{tabular}[c]{@{}c@{}}DLSI\end{tabular} & \begin{tabular}[c]{@{}c@{}}FDDL\end{tabular} & \begin{tabular}[c]{@{}c@{}}LRSDL\end{tabular} & \begin{tabular}[c]{@{}c@{}}SVM\end{tabular} & \begin{tabular}[c]{@{}c@{}}\mbox{K-SLD$^2$}\end{tabular} \\ \hline
\begin{tabular}[c]{@{}c@{}}Test sample  \\ classification accuracy (\%)\end{tabular}          & 79.36                                                       & 85.62                                                     & 88.43                                                      &88.14& 80.43 & \textbf{88.86}   \\ \hline
\begin{tabular}[c]{@{}c@{}}Number of \\parameters for representation\end{tabular} & 102400                                                       & 102400                                                    &
102400                                                    & $\sim$102400 &
$\sim$102400                                                    & \textbf{9600 }                                                       \\ \hline
\begin{tabular}[c]{@{}c@{}}Average training \\ time (sec)\end{tabular} & \textbf{0.034}                                                     & 1.2697                                                    &
9.3254                                                    & 50.5129 &
2.234                                                   & 0.111035                                                        \\ \hline
\begin{tabular}[c]{@{}c@{}}Normalized \\ reconstruction error\end{tabular} & 0.290                                                       & 0.363                                                    &
0.346                                                    &0.376 &
---                                                    & \textbf{0.178}                                                        \\ \hline
\end{tabular}
\caption{Comparison between different approaches for extended YaleB face recognition dataset.}
\label{table:result}
\end{table*} 

The best hyper-parameters for \mbox{K-SLD$^2$} turn out to be $n_1 = 13$, $n_2 = 17$, and $\mu = 0.9$. For standard subspace model, we obtain the best classification accuracy for $10$ dictionary atoms. The \mbox{K-SLD$^2$} uses more atoms overall, but each atom is described by fewer parameters. In Table \ref{table:result}, we compare the classification accuracy of \mbox{K-SLD$^2$} with the other dictionary learning methods. \mbox{K-SLD$^2$} offers better performance in this case, rather close to FDDL and correctly classify 11.16\% of the images than the baseline method. Furthermore, \mbox{K-SLD$^2$} learns a much more compact model, needing on the order of $1/10$th of the parameters of any other method.

We also calculate the \emph{normalized reconstruction error} (NRE) for all the learning algorithms as follows:
\begin{equation*}
\text{NRE} = \frac{||\mathbf{Y} - \mathbf{\hat{Y}}||^2}{||\mathbf{Y} ||^2},
\end{equation*}
where $\mathbf{Y}$ is the signal of interest and $\mathbf{\hat{Y}}$ is the reconstructed signal. Table \ref{table:result} shows that \mbox{K-SLD$^2$} provides the smallest NRE, reducing the error by 38.19\% over the baseline. Finally, we observe that the computational complexity, measured in training runtime on a standard desktop computer, is small. LRSDL method requires 50.51 seconds for training while \mbox{K-SLD$^2$} model requires only 0.11 seconds.
\begin{figure}[htbp]
\centering
\begin{subfigure}[b]{0.35\textwidth}
  \centering
  \includegraphics[width=\textwidth]{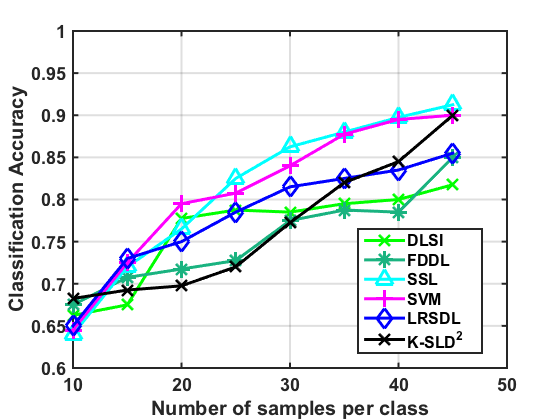}
  \caption{}
  \label{fig::EEGAccuracy}
\end{subfigure}
\begin{subfigure}[b]{0.35\textwidth}
  \centering
  \includegraphics[width=\textwidth]{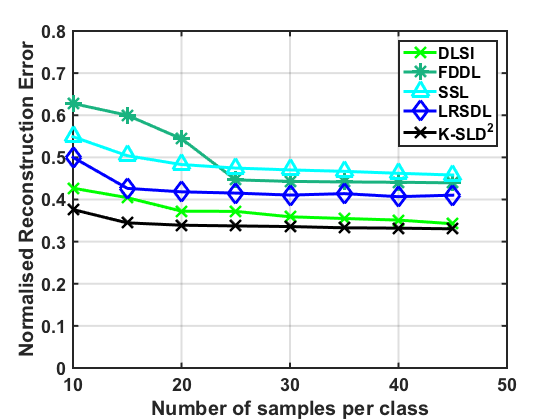}
  \caption{}
  \label{fig::EEGRecon}
\end{subfigure}
\caption{Performance on EEG Signal dataset (a) classification accuracy (b) normalized reconstruction error}
\label{fig::EEG_Performance}
\end{figure}

\begin{figure*}[htbp]
\centering
\begin{subfigure}[b]{0.32\textwidth}
  \centering
  \includegraphics[width=0.8\textwidth]{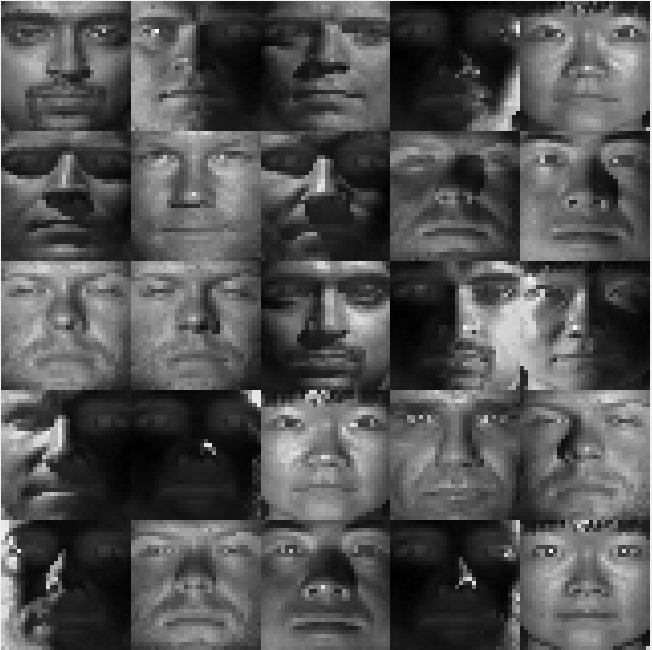}
  \caption{}
  \label{fig:sub1}
\end{subfigure}
~
\begin{subfigure}[b]{0.32\textwidth}
  \centering
  \includegraphics[width=0.8\textwidth]{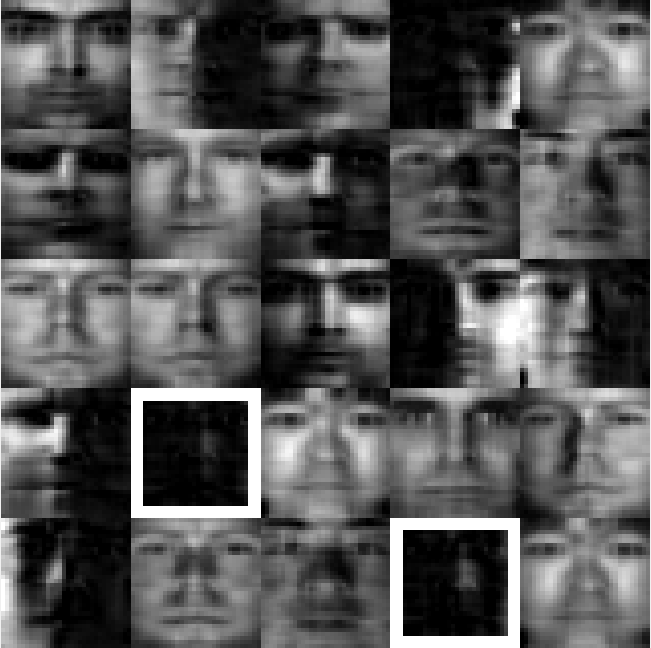}
  \caption{}
  \label{fig:sub2}
\end{subfigure}
~
\begin{subfigure}[b]{0.32\textwidth}
  \centering
  \includegraphics[width=0.8\textwidth]{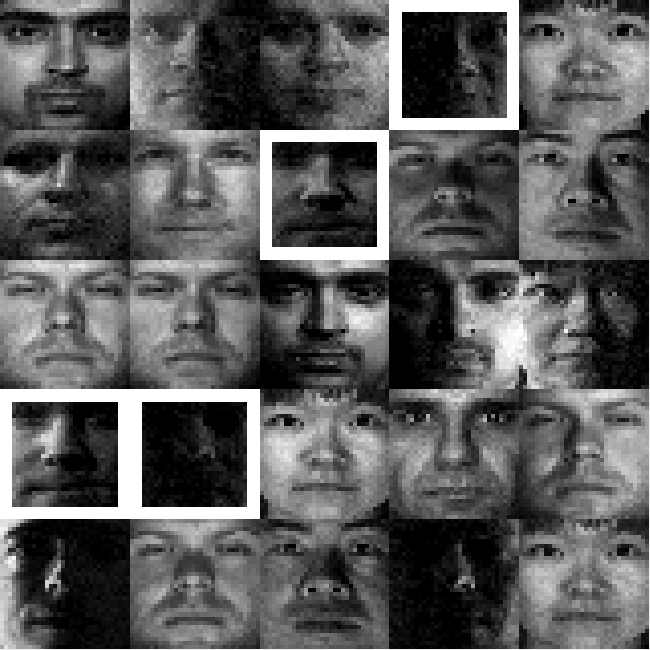}
  \caption{}
  \label{fig:sub3}
\end{subfigure}
  \caption{(a) A subset of test samples; (b) Image reconstruction and classification using K-S dictionary (\mbox{K-SLD$^2$}); (c) Image reconstruction and classification using standard subspace dictionary; \{White box indicates incorrect classification\}}
\label{fig:test}
\end{figure*}

We show  the classification and representation performance as a function of the size of the training set in Figs. \ref{fig::YaleAccuracy} and \ref{fig::YaleRecon}, respectively. When the number of samples for training is very small, say 10 samples per class, \mbox{K-SLD$^2$} model performance is superior, owing in part to the compact model. However, other methods outperform \mbox{K-SLD$^2$} as the number of samples increases.
%model prediction gets more accurate but less accurate than the other methods as can be seen in Fig. \ref{fig::YaleAccuracy}. We suspect that, from the slope of the \mbox{K-SLD$^2$} model prediction curve, with more training samples $(>50)$ \mbox{K-SLD$^2$} model prediction accuracy will surpass the other techniques. Because with 50 training samples per class the slope of prediction curve of all the other techniques is almost 0 while for \mbox{K-SLD$^2$} model slope is positive. 
On the other hand, the reconstruction error of \mbox{K-SLD$^2$} model is always smaller than other methods for any number of training samples as evident in Fig. \ref{fig::YaleRecon}. In Fig. \ref{fig:test}, we show a subset of raw YaleB face images used for the reconstruction and classification and compare the performance of \mbox{K-SLD$^2$} with SSL, where face in white box are the ones with the wrong label prediction. 

\subsubsection{EEG Dataset}
  We evaluate the performance of \mbox{K-SLD$^2$} on the UCI EEG dataset \cite{zhang1995event}, where EEG from the brain were recorded by placing the 64 electrodes on the scalp sampled at 256 Hz for 1 second to examine the correlation of EEG signal to an individual's alcoholism. Here, we obtain a 2-D signal with electrodes on one axis and the corresponding electrical signal time series on the other. This classification problem is analogous to binary classification having two categories of individuals either belongs to alcoholism or controlled group. The full datasets contains 120 trials for 122 subjects. Similar to YaleB face recognition dataset, we use 10 signals per class for training/dictionary learning and the remaining images for testing and find the value of $n_1 = 10$ and $n_2 = 6$ using cross-validation. 
\begin{table*}[ht]
\centering
\begin{tabular}{|c|c|c|c|c|c|c|}
\hline
  & \begin{tabular}[c]{@{}c@{}}SSL\end{tabular} & \begin{tabular}[c]{@{}c@{}}DLSI\end{tabular} & \begin{tabular}[c]{@{}c@{}}FDDL\end{tabular} & \begin{tabular}[c]{@{}c@{}}LRSDL\end{tabular} & \begin{tabular}[c]{@{}c@{}}SVM\end{tabular} & \begin{tabular}[c]{@{}c@{}}\mbox{K-SLD$^2$}\end{tabular} \\ \hline
\begin{tabular}[c]{@{}c@{}}Test sample  \\ classification accuracy (\%)\end{tabular}          & 64  & 66.25 & 67.5 & 65 & 64.9 & \textbf{68.25}   \\ \hline
\begin{tabular}[c]{@{}c@{}}Number of \\parameters for representation\end{tabular} & 163840& 163840  & 163840  &$\sim$ 163840 & $\sim$ 163840 & \textbf{2176}  \\ \hline
\begin{tabular}[c]{@{}c@{}}Average training \\ time (sec)\end{tabular} & 
\textbf{0.012} & 0.9546 & 3.4301 & 128.6921& 1.12 & 0.04 \\ \hline
\begin{tabular}[c]{@{}c@{}}Normalized \\ reconstruction error\end{tabular} & 
0.290  & 0.363  & 0.346  &0.50 & ---  & \textbf{0.178}\\ \hline
\end{tabular}
\caption{Comparison between different approaches on EEG signal dataset}
\label{table:EEGresult}
\end{table*} 

We compare the performance of \mbox{K-SLD$^2$} in Table \ref{table:EEGresult} with other dictionary learning methods. Again, \mbox{K-SLD$^2$} gives better classification performance and requires very few model parameters. In terms of NRE, \mbox{K-SLD$^2$} reconstruction error is less than 41\% of the best among the other methods. We obtain this performance gain for \mbox{K-SLD$^2$} because the dictionaries with separable structure are very good at signal representation \cite{hawe2013separable}. Similarly, we plot the classification and reconstruction accuracy in Figs. \ref{fig::EEGAccuracy} and \ref{fig::EEGRecon}, respectively. Here, by contrast to YaleB, we observe competitive performance for a larger number of training samples, due perhaps to the explicitly multidimensional nature of EEG signals. Reconstruction performance, measured in NRE, remains superior to other methods.

\section{Conclusion} 
We derive the performance limits on the classification performance of Kronecker-structured models. We derive an exact expression for the slope of misclassification probability as the noise power goes to zero. In high SNR regime, we derive a more accurate and tighter bound on misclassification probability which is determined by the product of principal angles between Kronecker subspaces.  We determine the upper and lower bounds on the rate at which the number of classes can grow as the signal dimension goes to infinity. We have also proposed a dictionary learning algorithm \mbox{K-SLD$^2$}, for fast classification and compact representation of multidimensional signals. This algorithm balances the learning of class-specific, Kronecker-structured subspaces against the learning of an general overcomplete dictionary that allows for the representation of general signals. Finally we show that \mbox{K-SLD$^2$} has improved classification performance over state-of-the-art dictionary learning methods, especially when the size of the training set is small, and competitive reconstruction performance in general.

% or


\appendices

\section{Diversity order gap}\label{Appendix:gap}
Given the K-S diversity order $d_{\text{K-S}}$ and the standard subspace diversity order $d_{\text{STD}}$. We derive the diversity gap
\begin{align*}
\gamma &= -[2n_1n_2-m_1m_2]^++[2n_1-m_1]^+[2n_2-m_2]^+,
\end{align*}
in terms of signal dimensions for different regions:\\
\textbf{Region 1:} $m_1 > 2n_1$ and $m_2>2n_2$\\
Since $m_1 > 2n_1$ and $m_2>2n_2$ therefore, ${[2n_1-m_1]^+ =}~ [2n_2-m_2]^+ = 0$ and $m_1m_2>4n_1n_2$ makes $[2n_1n_2-m_1m_2]^+ = 0$.\\
\textbf{Region 2:} $n_1<m_1<2n_1$ and $m_2>2n_2$\\
% \begin{align*}
% \gamma &= d_{\text{STD}} - d_{\text{K-S}}\\
% &= -[2n_1n_2-m_1m_2]^++[2n_1-m_1]^+[2n_2-m_2]^+\\
% & = 0.
% \end{align*}
Since $m_2>2n_2$ therefore, $[2n_2-m_2]^+ = 0$ also $m_1>n_1$ and $m_2>2n_2$ implies $m_1m_2>2n_1n_2$ therefore, ~$[2n_1n_2-~m_1m_2]^+ = ~0$.\\
\textbf{Region 3:} $n_2<m_2<2n_2$ and $m_1>2n_1$\\
Using the similar argument the diversity gap is 0.\\
\textbf{Region 4:} $n_1<m_1<2n_1$ and $n_2<m_2<2n_2$\\
% \begin{align*}
% \gamma &= d_{\text{STD}} - d_{\text{K-S}}\\
% &= -[2n_1n_2-m_1m_2]^++[2n_1-m_1]^+[2n_2-m_2]^+.
% \end{align*}
Since $n_1<m_1<2n_1$ and $n_2<m_2<2n_2$ implies that $n_1n_2<m_1m_2<4n_1n_2$, this gives rise to two different subregions which are $m_1m_2>2n_1n_2$ and $m_1m_2<2n_1n_2$.\\
For $m_1m_2>2n_1n_2$, $n_1<m_1<2n_1$ and $n_2<m_2<2n_2$ which implies $[2n_1n_2-m_1m_2]^+ = 0$ we derive the diversity order gap as:
\begin{align*}
\gamma &= -[2n_1n_2-m_1m_2]^++[2n_1-m_1]^+[2n_2-m_2]^+\\
& = (2n_1-m_1)(2n_2-m_2).
\end{align*}
On the other hand if $m_1m_2<2n_1n_2$, $n_1<m_1<2n_1$ and $n_2<m_2<2n_2$,
\begin{align*}
\gamma &= -[2n_1n_2-m_1m_2]^++[2n_1-m_1]^+[2n_2-m_2]^+\\
& = -(2n_1n_2-m_1m_2)+(2n_1-m_1)(2n_2-m_2)\\
&=2(m_1-n_1)(m_2-n_2).
\end{align*}{}

\section{Intersection of Kronecker Subspaces}\label{Appendix:DimInter}
Here we characterize the dimension of intersections of subspaces spanned by Kronecker products of matrices. To the best of our knowledge this result is not in the literature, although its statement is intuitive.
\begin{lemma} 
\label{lemma::intersectionSpaces}
	Suppose $\dim[\mathcal{R}(A_i) \bigcap \mathcal{R}(A_j)]~=~x$ and $\dim[\mathcal{R}(B_i) \bigcap \mathcal{R}(B_j)]~=~y$, where $\mathcal{R}(\cdot)$ denotes the range space of a matrix. Then,
	\begin{equation}
		\dim[\mathcal{R}(\mathbf{B}_i \otimes \mathbf{A}_i) \bigcap \mathcal{R}(\mathbf{B}_j \otimes \mathbf{A}_j)] = xy.
	\end{equation}
\end{lemma}
\begin{IEEEproof}
	From \cite[p. 447]{bernstein2009matrix} for $p \in \mathbb{R}^{m_1 \times n_1} $ and $q \in \mathbb{R}^{m_2 \times n_2}$, we have
	\begin{equation}
		\mathcal{R}(p\otimes q) = \mathcal{R}(p\otimes \mathbf{I}_{m_2\times m_2}) \bigcap \mathcal{R}(\mathbf{I}_{m_1\times m_1} \otimes q).
	\end{equation}
	Therefore, we can write the dimension as
	\begin{multline}
 		\dim\left[\mathcal{R}(\mathbf{B}_i \otimes \mathbf{A}_i) \bigcap \mathcal{R}(\mathbf{B}_j \otimes \mathbf{A}_j)\right] =\\
		\dim \Big[\mathcal{R}(\mathbf{B}_i \otimes \mathbf{I}_{m_1\times m_1}) \bigcap \mathcal{R}(\mathbf{I}_{m_2\times m_2} \otimes \mathbf{A}_i) \bigcap\\   \mathcal{R}(\mathbf{B}_j \otimes \mathbf{I}_{m_1\times m_1}) \bigcap \mathcal{R}(\mathbf{I}_{m_2\times m_2} \otimes \mathbf{A}_j)\Big].
	\end{multline}
	Rearranging terms, we obtain
	%\begin{equation}
	\begin{multline}
	\label{eq::model::1}
		\dim\left[\mathcal{R}(\mathbf{B}_i \otimes \mathbf{A}_i) \bigcap \mathcal{R}(\mathbf{B}_j \otimes \mathbf{A}_j)\right] = \\ \dim\Big[\left[\mathcal{R}(\mathbf{B}_i \otimes \mathbf{I}_{m_1\times m_1}) \bigcap  \mathcal{R}(\mathbf{B}_j \otimes \mathbf{I}_{m_1\times m_1})\right] \\ \bigcap  \left[\mathcal{R}(\mathbf{I}_{m_2\times m_2} \otimes \mathbf{A}_i) \bigcap \mathcal{R}(\mathbf{I}_{m_2\times m_2} \otimes \mathbf{A}_j)\right]\Big].
	\end{multline}

	Next, let $\mathbf{A}_{ij}$ and $\mathbf{B}_{ij}$ be matrices whose column spans are $\mathcal{R}(\mathbf{A}_i) \bigcap  \mathcal{R}(\mathbf{A}_j)$ and $\mathcal{R}(\mathbf{B}_i) \bigcap \mathcal{R}(\mathbf{B}_j)$, respectively. It is straightforward to verify that
	\begin{equation}
		\mathcal{R}(\mathbf{B}_i \otimes \mathbf{I}_{m_1\times m_1}) \bigcap  \mathcal{R}(\mathbf{B}_j \otimes \mathbf{I}_{m_1\times m_1}) = 
		\mathcal{R}(\mathbf{B}_{ij} \otimes \mathbf{I}_{m_1\times m_1}),
	\end{equation}
	and
	\begin{equation}
		\mathcal{R}(\mathbf{I}_{m_2\times m_2} \otimes \mathbf{A}_i) \bigcap \mathcal{R}(\mathbf{I}_{m_2\times m_2} \otimes \mathbf{A}_j) = 
		\mathcal{R}(\mathbf{I}_{m_2\times m_2} \otimes \mathbf{A}_{ij}).
	\end{equation}
	Therefore, we can rewrite the subspace dimension as
	\begin{multline}
		\dim\left[\mathcal{R}(\mathbf{B}_i \otimes \mathbf{A}_i) \bigcap \mathcal{R}(\mathbf{B}_j \otimes \mathbf{A}_j)\right] = \\
		\dim\left[\mathcal{R}( \mathbf{B}_{ij} \otimes \mathbf{I}_{m_1\times m_1})\bigcap \mathcal{R}(\mathbf{I}_{m_2\times m_2} \otimes \mathbf{A}_{ij} )\right].
	\label{eq::model::3}
	\end{multline}
	Next, we can apply the lemma of \cite[p. 447]{bernstein2009matrix} in reverse, yielding
	\begin{align*}
		\dim[\mathcal{R}(\mathbf{B}_i \otimes \mathbf{A}_i) \bigcap \mathcal{R}(\mathbf{B}_j \otimes \mathbf{A}_j)] &=  \dim[\mathcal{R}( \mathbf{B}_{ij} \otimes  \mathbf{A}_{ij} )]\\
		&=  r(\mathbf{B}_{ij}) \cdot r(\mathbf{A}_{ij})\\
		&=  x y.
		% \label{eq::model::4}
	\end{align*}

\end{IEEEproof}

\section{Proof of Theorem \ref{thm:Geometry}} \label{Appendix:Geometry}
Expanding the Bhattacharyya bound from \eqref{eq::Bhattacharya} we obtain the misclassification probability bound in terms of $\lambda_{ij}$ the nonzero eigenvalues of $\mathbf{D}_i\mathbf{D}_i^T + \mathbf{D}_j\mathbf{D}_j^T $ in \eqref{eq::CCB::14} as:
\begin{align}
P_e(\mathbf{D}_i,\mathbf{D}_j) &\leq \frac{1}{2} \left(\frac{1}{\sigma^2}\right)^{-\frac{r^*_{ij}-n_1n_2}{2}} \nonumber \\& \, \cdot \left( \frac{\prod_{l=1}^{r^*_{ij}}(\lambda_{ijl}+\sigma^2)}{\sqrt{\prod_{l=1}^{n_1n_2}(\lambda_{il} + \sigma^2) \cdot \prod_{l=1}^{n_1n_2}(\lambda_{jl}+\sigma^2)}} \right)^{-\frac{1}{2}} \nonumber \\
&= 2^{\frac{n_1n_2-2}{2}} \cdot \left(\frac{1}{\sigma^2}\right)^{-\frac{r^*_{ij}-n_1n_2}{2}}\nonumber \\ & \, \cdot \left( \frac{\prod_{l=1}^{r^*_{ij}}\lambda_{ijl}}{\sqrt{\prod_{l=1}^{n_1n_2}\lambda_{il} \cdot \prod_{l=1}^{n_1n_2}\lambda_{jl}}} \right)^{-\frac{1}{2}} + o ((\sigma^2)^{\frac{r^*_{ij}-n_1n_2}{2}})
\label{eq::ApB::1}
\end{align}
Now, our aim is to expand $\prod_{l=1}^{r^*_{ij}}\lambda_{ijl}$ in terms of principal angles.
\begin{multline*}
\prod_{l=1}^{r^*_{ij}}\lambda_{ijl} = \mathrm{pdet}(U_{i,\cap}\lambda_{i,\cap} U_{i,\cap}^T + U_{j,\cap}\lambda_{j,\cap} U_{j,\cap}^T + \\  U_{i,\setminus}\lambda_{i,\setminus} U_{i,\setminus}^T \nonumber + U_{j,\setminus}\lambda_{j,\setminus} U_{j,\setminus}^T).
\end{multline*}
As the image of $U_{i,\cap}$ is orthogonal to $U_{i,\setminus}$ we can write:
\begin{multline*}
\prod_{l=1}^{r^*_{ij}}\lambda_{ijl} = \mathrm{pdet}(U_{i,\cap}\lambda_{i,\cap} U_{i,\cap}^T + U_{j,\cap}\lambda_{j,\cap} U_{j,\cap}^T) \times \\ \mathrm{pdet}(U_{i,\setminus}\lambda_{i,\setminus} U_{i,\setminus}^T + U_{j,\setminus}\lambda_{j,\setminus} U_{j,\setminus}^T).
\end{multline*}
Following few simple mathematical steps as described in \cite{huang2016role} we obtain:
\begin{multline*}
\prod_{l=1}^{r^*_{ij}}\lambda_{ijl} =  \mathrm{pdet}(U_{i,\cap}\lambda_{i,\cap} U_{i,\cap}^T + U_{j,\cap}\lambda_{j,\cap} U_{j,\cap}^T) \cdot \mathrm{det}(\lambda_{i,\setminus}) \times \\ \mathrm{det}(\lambda_{j,\setminus}^{\frac{1}{2}}(\mathbf{I} -  U_{j,\setminus}^T U_{i,\setminus} U_{i,\setminus}^T U_{j,\setminus})\lambda_{j,\setminus}^{\frac{1}{2}}).
\end{multline*}
By expanding $U_{i,\cap}, U_{j,\cap}, U_{i,\setminus}, U_{j,\setminus}$ in terms of their row and columns subspace Kronecker products and then following some simple Kronecker product properties we obtain:
\begin{multline*}
\prod_{l=1}^{r^*_{ij}}\lambda_{ijl} = \mathrm{pdet}(U_{i,\cap}\lambda_{i,\cap} U_{i,\cap}^T + U_{j,\cap}\lambda_{j,\cap} U_{j,\cap}^T) \cdot \mathrm{det}(\lambda_{i,\setminus}) \times \\  \mathrm{det}(\lambda_{j,\setminus}^{\frac{1}{2}}(\mathbb{I} -  (U_{j,\setminus}^{AT} U_{i,\setminus}^A U_{i,\setminus}^{AT} U_{j,\setminus}^A) \otimes (U_{j,\setminus}^{BT} U_{i,\setminus}^B U_{i,\setminus}^{BT} U_{j,\setminus}^B))\lambda_{j,\setminus}^{\frac{1}{2}}).
\end{multline*}
By careful inspection of $\mathrm{det}(U_{j,\setminus}^{AT} U_{i,\setminus}^A U_{i,\setminus}^{AT} U_{j,\setminus}^A)$ we find that product of eigenvalues of $(U_{j,\setminus}^{AT} U_{i,\setminus}^A) (U_{j,\setminus}^{AT} U_{i,\setminus}^{A} )^T$ is the square of the singular values of $(U_{j,\setminus}^{AT} U_{i,\setminus}^A)$ and are the cosines square of the principal angles between the subspaces. Therefore we obtain:
\begin{multline*}
\prod_{l=1}^{r^*_{ij}}\lambda_{ijl} = \mathrm{pdet}(U_{i,\cap}\lambda_{i,\cap} U_{i,\cap}^T + U_{j,\cap}\lambda_{j,\cap} U_{j,\cap}^T) \times \\ \prod_{l=1}^{n_1n_2-r_{\cap}}\lambda_{i,\setminus,l}\cdot \prod_{l=1}^{n_1n_2-r_{\cap}}\lambda_{j,\setminus,l} \cdot \prod_{l=r_{\cap}+1}^{n_1n_2}(1-\cos^2(\theta_l)).
\end{multline*}
In terms of principal angles of individual row and column subspaces we obtain:
\begin{multline*}
\prod_{l=1}^{r^*_{ij}}\lambda_{ijl} = \mathrm{pdet}(U_{i,\cap}\lambda_{i,\cap} U_{i,\cap}^T + U_{j,\cap}\lambda_{j,\cap} U_{j,\cap}^T) \times \\ \prod_{l=1}^{n_1n_2-r_{\cap}}\lambda_{i,\setminus,l} \prod_{l=1}^{n_1n_2-r_{\cap}}\lambda_{j,\setminus,l} \prod_{l=t_1+1}^{n_1}\prod_{l=t_2+1}^{n_2}(1-\cos^2(\theta_l^A)\cos^2(\theta_l^B)).
% \label{eq::ApB::2}
\end{multline*}
Substituting this in \eqref{eq::ApB::1}, we obtain the desired results as stated in Theorem \ref{thm:Geometry}.

% \section*{Acknowledgment}


\ifCLASSOPTIONcaptionsoff
  \newpage
\fi

\bibliographystyle{IEEEtran}

\bibliography{ref}


% \end{IEEEbiography}


% % insert where needed to balance the two columns on the last page with
% %\newpage



%\vfill



\end{document}